\documentclass[journal=jacsat,manuscript=article, natbib=true]{achemso}
\setcitestyle{numbers,square}
\usepackage[version=3]{mhchem}
\usepackage{adjustbox}
\usepackage{threeparttable}
\usepackage{booktabs}
\usepackage[font=bf]{caption}
\usepackage{graphicx}
\usepackage{subcaption}  
\usepackage{rotating}
\usepackage{bm}
\usepackage{xcolor}
\usepackage[normalem]{ulem}

\usepackage{amsmath,amssymb,mathtools,bm}
\usepackage{braket}
\usepackage{booktabs,multirow,array}
\usepackage{siunitx}
\usepackage{xcolor}

\SectionNumbersOn

\renewcommand{\dag}{^{\dagger}}
\newcommand{\cre}[1]{\hat a^{\dagger}_{#1}}
\newcommand{\ann}[1]{\hat a_{#1}}
\newcommand{\bc}{\hat b^{\dagger}}
\newcommand{\ba}{\hat b}
\newcommand{\dt}{\tilde d}
\newcommand{\lam}{\bm{\lambda}}
\newcommand{\asym}[4]{\langle #1#2\Vert #3#4\rangle}
\newcommand{\asyl}[4]{\langle #1#2\Vert #3#4\rangle_{\lambda}}
\newcommand{\Hop}{\hat H}
\newcommand{\cc}{^{*}}
\newcommand{\sq}{\sqrt{\frac{\omega}{2}}}

\newcommand{\M}{\mathbf M}
\newcommand{\corr}[1]{\textcolor{red}{#1}}   

\usepackage[colorlinks=true,
            linkcolor=blue,
            citecolor=magenta,
            urlcolor=blue]{hyperref}

\author{Sudipta Chakraborty}
\affiliation{Department of Chemistry, Indian Institute of Technology Bombay, Mumbai 400076, India}

\author{Amrita Manna}
\affiliation{Department of Chemistry, Indian Institute of Technology Bombay, Mumbai 400076, India}
\author{Lukas Konecny}
\affiliation{Institute of Inorganic Chemistry, Slovak Academy of Sciences, 84536 Bratislava, Slovakia}
\alsoaffiliation{Department of Inorganic Chemistry, Faculty of Natural Sciences, Comenius University, 84215 Bratislava, Slovakia}

\author{Achintya Kumar Dutta}
\email{achintya@chem.iitb.ac.in}
\affiliation{Department of Chemistry, Indian Institute of Technology Bombay, Mumbai 400076, India}

\title[An \textsf{achemso} demo]
  {Relativistic Cavity Quantum Electrodynamical Algebraic Diagrammatic Construction Method for Strong Light–Matter Coupling}

\abbreviations{IR, NMR, UV}

\begin{document}

\newpage

\begin{abstract}

We present a relativistic four-component implementation of cavity quantum electrodynamics at the level of the second-order algebraic diagrammatic construction, 4c-QED-ADC(2). Starting from the origin-invariant Pauli--\allowbreak Fierz Hamiltonian in the coherent-state basis, we include the cavity either self-consistently in a 4c-QED Dirac--\allowbreak Hartree--\allowbreak Fock reference or only at the correlated level on a cavity-free reference.
We apply the method to systems with heavy elements, where spin–orbit coupling makes singlet–triplet transitions weakly allowed, an effect that a non-relativistic treatment cannot capture. We show that these spin-forbidden transitions form polaritons inside a cavity, and that the polaritons borrow intensity from bright singlet states and become up to an order of magnitude brighter than the free transition. The method is not limited to atoms and linear molecules, as shown for selenophene, where the weakly allowed triplet state forms polaritons that are about eleven times brighter than in the free molecule. For HBr, the cavity creates a bound well on an otherwise repulsive excited-state surface, and at stronger coupling a triplet state also couples to the cavity through spin–orbit coupling. Including the photonic degrees of freedom makes the matrix–vector product 2 and 3 times more costly compared to standard 4c-ADC(2) for one and two photons, respectively. Comparing the two references, we find that including the cavity only at the correlated level works well at weak coupling, but the differences grow with the coupling strength, so a cavity-relaxed 4c-QED-DHF reference is needed at strong coupling.
\end{abstract}
\maketitle

\newpage

\section{Introduction:}
Advances in optical microcavities, plasmonic structures, and nanophotonic platforms have opened up regimes in which the light-matter interaction strength becomes comparable to intrinsic electronic energy scales. When a molecular or condensed-matter system is placed inside an optical cavity, the electromagnetic vacuum modes are modified by the cavity, leading to a substantial enhancement of the light-matter coupling strength relative to free space. If a cavity resonance is tuned close to an internal excitation of the material, the system can no longer be described as independent photons and matter degrees of freedom. Instead, hybrid states, polaritons emerge, reflecting the coherent mixing of light with electronic or vibrational excitations.\cite{ruggenthaler2018quantum, frisk2019ultrastrong} The character of a polaritonic state is determined by the underlying material excitation, resulting in exciton polaritons, vibrational polaritons, or more complex hybrid states that combine multiple degrees of freedom. The ability to controllably engineer such hybridization has given rise to the rapidly expanding field of cavity quantum electrodynamical materials, which unites concepts from quantum optics, condensed matter physics, chemistry, and materials science. This interdisciplinary area encompasses polaritonic chemistry,\cite{ebbesen2016hybrid,ruggenthaler2018quantum, li2020origin, sidler2022perspective, mandal2023theoretical, simpkins2023control} cavity-modified electronic structure, light-induced phases of matter, and the manipulation of quantum states using both classical and quantized electromagnetic fields. Strong light-matter coupling has been shown to profoundly alter physical and chemical behavior. Experimental and theoretical studies have demonstrated cavity-induced modifications of chemical reaction pathways\cite{hutchison2012modifying,thomas2016cyriaque, garcia2021manipulating,schafer2022shining} and ground-state properties,\cite{flick2018ab,latini2021ferroelectric} enhanced energy and charge transport,\cite{schachenmayer2015cavity,zhong2016non,feist2015extraordinary} selective control over electronic excitations,\cite{stranius2018selective} and changes in electron-phonon interactions that can influence collective phenomena such as superconductivity.\cite{sentef2018cavity,curtis2019cavity, schlawin2019cavity,allocca2019cavity,li2020manipulating,andolina2024amperean}
For example, the coupling between photons and phonons can renormalize both the electron–phonon interaction and the phonon frequencies,\cite{sentef2018cavity} or drive the system into nonequilibrium states via the quantum Eliashberg mechanism.\cite{curtis2019cavity} In the context of excitonic systems, strong coupling enables control over exciton dynamics and has even led to the observation of exciton-polariton condensation.\cite{latini2019cavity,levinsen2019microscopic,forg2019cavity,keeling2020bose} More recently, quantum Hall systems placed inside cavities have revealed signatures of ultrastrong coupling, accompanied by substantial changes in their transport characteristics.\cite{paravicini2019magneto} At the same time, other studies are exploring structured and chiral electromagnetic environments, showing the growing variety of cavity-mediated phenomena.\cite{petersen2014chiral,zhang2019chiral,hubener2021engineering} A quantitative understanding of these effects requires an explicit quantum mechanical treatment of the electromagnetic field, as the interaction with vacuum fluctuations is central to cavity-induced phenomena. While simplified cavity-QED models such as the Jaynes-Cummings\cite{jaynescomparison63} or Dicke\cite{Dicke54} Hamiltonians have provided invaluable conceptual insights, they rely on severe truncations of the matter degrees of freedom. For realistic molecular and material systems, where subtle changes in electronic structure and spectroscopy are of interest, first-principles approaches that treat both electrons and photons on equal footing become essential.

These advances have inspired a parallel surge in theoretical development.\cite{mandal2023theoretical} Yet, the theoretical modeling of hybrid light-matter systems remains a deeply challenging endeavor.\cite{ruggenthaler2018quantum} These fundamentally new quantum states significantly elevate both the chemical and computational complexity by introducing a high-dimensional space that encompasses the coupled light and matter degrees of freedom. To address this complexity, a wide array of computational approaches has been proposed, ranging from quantum-optical models and semiclassical approximations\cite{li2020cavity} to fully \textit{ab initio} quantum electrodynamics (QED) treatments.\cite{riso2022molecular, ruggenthaler2014quantum, haugland2020coupled, bauer2023perturbation} Recent literature has focused on incorporating photonic degrees of freedom directly into ab initio electronic-structure methods, enabling a fully self-consistent description of light–matter interactions. Within this paradigm, several theoretical developments have emerged, including quantum electrodynamical extensions of density functional theory (QEDFT, also commonly referred to as QED-DFT),\cite{ruggenthaler2011time,ruggenthaler2014quantum,pellegrini2014optimized,flick2018ab,jestadt2019light,flick2020ab, vu2022enhanced, pavovsevic2022wavefunction, liebenthal2023assessing} real-time\cite{ruggenthaler2011time, PhysRevLett.110.233001, flick2017atoms, malave2022real} and linear-response formulations of QED-TDDFT,\cite{yang2021quantum, yang2022cavity} configuration-interaction approaches such as QED-CIS,\cite{mctague2022non} and cavity-adapted version of second-order the algebraic diagrammatic construction theory.\cite{bauer2023perturbation} Additional contributions span QED-coupled-cluster theories,\cite{haugland2020coupled, mordovina2020polaritonic, deprince2021cavity} variational two-electron reduced density-matrix frameworks,\cite{mallory2022reduced}, complete active space configuration interaction theory\cite{vu2024cavity} and even diffusion Monte Carlo treatments tailored for strongly coupled light-matter systems.\cite{weight2024diffusion} 

However, the vast majority of existing \textit{ab initio} cavity-QED (CQED) implementations rely on a non-relativistic Hamiltonian. This limitation becomes critical for systems containing heavy elements or when spin-dependent phenomena play a decisive role. Many well-established physical and chemical effects-ranging from characteristic trends across the periodic table to spectroscopic observables such as x-ray absorption, magnetic resonance, and phosphorescence are governed by relativistic effects that are only properly captured within a Dirac-equation-based framework.\cite{dirac28, dirac1928quantum, pyykko2012relativistic} In relativistic quantum chemistry, the four-component Dirac-Coulomb Hamiltonian represents the most rigorous standard, as it incorporates relativistic kinetic energy and spin-orbit coupling variationally and without reliance on perturbative approximations.\cite{saue2011relativistic} Extending cavity-QED methodologies to a fully relativistic framework is therefore essential for an accurate description of light-matter coupling in systems containing heavy elements. In particular, a four-component treatment provides direct access to spin-orbit-induced singlet-triplet mixing and spin-forbidden transitions, which are inaccessible or poorly described in non-relativistic theories. This capability is especially important for heavy element systems, where higher-order relativistic effects become prominent and strongly influence excited-state dynamics. Despite significant progress in non-relativistic CQED electronic-structure methods, relativistic CQED implementations remain notably limited. For instance, Konecny and coworkers have reported relativistic QED-DFT formulations employing both four-component (4c) and two-component (2c) Hamiltonians,\cite{Konecnyrelativistic2025,konecny2025linear} and Thiam et al.\cite{thiam2025comprehensive} have recently introduced a 4c CQED Dirac-Hartree-Fock (4c-CQED-DHF) scheme. In this work, we advance the theoretical description of cavity-coupled systems by implementing a 4c CQED extension of the algebraic diagrammatic construction method at second order, denoted 4c-QED-EE-ADC(2). Section 2 outlines the theoretical framework for constructing the reference state using the relativistic four-component CQED Dirac-Hartree-Fock (4c-QED-DHF) approach, along with the formulation of the 4c-QED-EE-ADC(2) method. We also describe an alternative implementation that employs a non-polaritonic 4c-DHF reference, wherein CQED effects are incorporated exclusively at the electron correlation level. Section 4 then presents numerical results and analysis from 4c-QED-EE-ADC(2) calculations for molecular systems containing heavy elements.

\section{Theory}

\subsection{Relativistic Hartree-Fock using Pauli-Fierz Hamiltonian}
The electronic Hamiltonian is the
no-pair Dirac--Coulomb Hamiltonian\cite{dyall2007introduction}
\begin{equation}
  \Hop_{\mathrm{DC}}^{+} = \hat\Lambda^{+}\Bigl[\sum_{i=1}^{N} \hat h_{\mathrm D}(i)
     + \sum_{i<j}\frac{1}{r_{ij}}\Bigr]\hat\Lambda^{+},
  \qquad
  \hat h_{\mathrm D} = c\,\bm\alpha\cdot\hat{\mathbf p} + (\beta - I_4)c^2 + V_{\mathrm{nuc}}(\mathbf r),
  \label{eq:hdc}
\end{equation}
where $\bm\alpha$ and $\beta$ are the Dirac matrices, $c$ is the speed of
light, $V_{\mathrm{nuc}}$ the nuclear potential, and $\hat\Lambda^{+}$ the
projector onto the electronic (positive-energy) states of the
mean-field Fock operator.
The molecule interacts with a single cavity mode of frequency $\omega$ and
polarization $\mathbf e$, described by the creation and annihilation
operators $\bc$ and $\ba$. In the long-wavelength approximation, the vector
potential of the mode, $\hat{\mathbf A}(0)$, is uniform over the molecule.
Because $\hat h_{\mathrm D}$ depends on $\hat{\mathbf A}$ only through the
linear term $c\,\bm\alpha\cdot(\hat{\mathbf p}+\hat{\mathbf A}/c)$, the
unitary length-gauge (Power--Zienau--Woolley) transformation
$\hat U=\exp[\mathrm i\,\hat{\mathbf A}(0)\cdot\hat{\bm\mu}_{\mathrm e}/c]$,
with $\hat{\bm\mu}_{\mathrm e}$ the electronic dipole operator, removes the
minimal coupling exactly, as in the non-relativistic case, and generates
the bilinear and dipole self-energy (DSE) terms.\cite{konecny2025linear,Konecnyrelativistic2025,thiam2025comprehensive} The length-gauge Hamiltonian
of the molecule in the cavity is therefore
\begin{equation}
  \Hop = \Hop_{\mathrm{DC}}^{+} + \omega\,\bc\ba
       - \sq\,(\lam\cdot\hat{\bm\mu})\,(\bc+\ba)
       + \tfrac12(\lam\cdot\hat{\bm\mu})^{2}.
  \label{eq:pf}
\end{equation}
Here $\hat{\bm\mu}=-\sum_i\mathbf r_i+\sum_A Z_A\mathbf R_A$ is the
molecular dipole operator, and $\lam=\lambda\,\mathbf e$ is the coupling
vector, where $\lambda$ is called coupling strength. The last
two terms of Eq~\ref{eq:pf} are the bilinear coupling and the DSE. In the
four-component space, $\mathbf r$ multiplies the $4\times4$ unit matrix, so
the dipole couples large to large and small to small components only. 
The coherent-state transformation\cite{haugland2020coupled} with the transformation operator
$\hat U(z)=\exp[z(\bc-\ba)]$ gives
\begin{equation}
  \hat U\dag\Hop\hat U = \Hop_{\mathrm{DC}}^{+} + \omega\,\bc\ba
     - \sq\,\bigl[\lam\cdot(\hat{\bm\mu}-\braket{\hat{\bm\mu}})\bigr](\bc+\ba)
     + \tfrac12\bigl[\lam\cdot(\hat{\bm\mu}-\braket{\hat{\bm\mu}})\bigr]^{2}.
\end{equation}
The nuclear dipole cancels in $\hat{\bm\mu}-\braket{\hat{\bm\mu}}$ and we write
$\hat d=\lam\cdot\sum_i\mathbf r_i$, so that
$\lam\cdot(\hat{\bm\mu}-\braket{\hat{\bm\mu}})=-(\hat d-\braket{\hat d})$.
A photon-parity transformation with
$\hat\Pi=\mathrm e^{\mathrm i\pi\bc\ba}$ then gives the Hamiltonian used
in the rest of this work,
\begin{equation}
  \
  \Hop = \Hop_{\mathrm{DC}}^{+} + \omega\,\bc\ba
       - \sq\,\bigl(\hat d-\braket{\hat d}\bigr)(\bc+\ba)
       + \tfrac12\bigl(\hat d-\braket{\hat d}\bigr)^{2}\
  \label{eq:hcs}
\end{equation}
Equation~\ref{eq:hcs} is therefore origin invariant for neutral and charged
systems alike, and the mean-field energy derived from it does not depend on $\omega$.\cite{haugland2020coupled,foley2023ab}

We expand in an orthonormal set of positive-energy four-component spinors
$\{\psi_p\}$ (indices $i,j,k,l$ denotes occupied, $a,b,c,d$ denotes virtual, $p,q,r,s$ denotes general). Their matrix elements of $\hat d$ contain large- and
small-component contributions,
\begin{equation}
  d_{pq} = \braket{\psi_p|\lam\cdot\mathbf r|\psi_q}
         = \lam\cdot\bigl(\braket{\psi_p^{\mathrm L}|\mathbf r|\psi_q^{\mathrm L}}
           + \braket{\psi_p^{\mathrm S}|\mathbf r|\psi_q^{\mathrm S}}\bigr),
  \qquad d_{qp}=d_{pq}\cc .
\end{equation}
On the $N$-electron space, $\hat d-\braket{\hat d}$ is the one-body
operator with matrix
\begin{equation}
  \dt_{pq} = d_{pq} - \bar d\,\delta_{pq},
  \qquad \bar d = \frac{\braket{\hat d}}{N} = \frac1N\sum_{i}d_{ii},
  \label{eq:dtilde}
\end{equation}
where the sum runs over all $N$ occupied spinors. We call $\dt$ the coherent-state dipole. It is origin invariant,
and its occupied trace vanishes, $\sum_i\dt_{ii}=0$. The off-diagonal blocks
are unchanged, $\dt_{ia}=d_{ia}$. The DSE then becomes
\begin{equation}
  \tfrac12(\hat d-\braket{\hat d})^{2}
   = \tfrac12\sum_{pq}\tilde Q_{pq}\,\cre p\ann q
   + \tfrac14\sum_{pqrs} D_{pqrs}\,\cre p\cre q\ann s\ann r,
  \qquad
  D_{pqrs} = \dt_{pr}\dt_{qs}-\dt_{ps}\dt_{qr},
  \label{eq:dse2q}
\end{equation}
where $\tilde Q_{pq}=\braket{\psi_p|(\lam\cdot\mathbf r-\bar d)^2|\psi_q}$
is the one-electron quadrupole-type operator. The two-electron part of the
DSE has the same permutational symmetry as the antisymmetrized
Dirac--Coulomb integrals,
$\asym pqrs=\braket{pq|rs}-\braket{pq|sr}$ and, Coulomb and DSE two-electron tensors can be combined into the DSE-dressed antisymmetrized integrals
\begin{equation}
  \asyl pqrs = \asym pqrs + \dt_{pr}\dt_{qs} - \dt_{ps}\dt_{qr}.
  \label{eq:dressed}
\end{equation}

The large and small components are expanded in two-spinor basis functions
$\{\chi_\mu\}$ and $\{(\bm\sigma\cdot\hat{\mathbf p})\chi_\mu/2c\}$ following restricted kinetic balance,
\begin{equation}
  \psi_p = \begin{pmatrix}\sum_\mu C^{\mathrm L}_{\mu p}\,\chi_\mu\\[2pt]
           \sum_\mu C^{\mathrm S}_{\mu p}\,\frac{1}{2c}(\bm\sigma\cdot\hat{\mathbf p})\chi_\mu\end{pmatrix}.
\end{equation}
with $S_{\mu\nu}=\braket{\chi_\mu|\chi_\nu}$,
$T_{\mu\nu}=\tfrac12\braket{\chi_\mu|\hat p^2|\chi_\nu}$,
$V_{\mu\nu}=\braket{\chi_\mu|V_{\mathrm{nuc}}|\chi_\nu}$, and
$W_{\mu\nu}=\braket{\bm\sigma\cdot\hat{\mathbf p}\chi_\mu|V_{\mathrm{nuc}}|\bm\sigma\cdot\hat{\mathbf p}\chi_\nu}$,
the one-electron Dirac operator and the metric are\cite{dyall2007introduction}
\begin{equation}
  \mathbf h = \begin{pmatrix} \mathbf V & \mathbf T\\ \mathbf T & \frac{1}{4c^2}\mathbf W-\mathbf T\end{pmatrix},
  \qquad
  \mathbf S = \begin{pmatrix} \mathbf S & \mathbf 0\\ \mathbf 0 & \frac{1}{2c^2}\mathbf T\end{pmatrix}.
\end{equation}
The 4c density matrix $\mathbf D=\mathbf C_{\mathrm{occ}}\mathbf C_{\mathrm{occ}}\dag$
has blocks $\mathbf D^{XY}$ ($X,Y\in\{\mathrm L,\mathrm S\}$), with elements $D^{XY}_{\tau\kappa}=\sum_i C^X_{\tau i}C^{Y*}_{\kappa i}$. The 4c-QED-DHF Fock matrix can be expressed as
\begin{subequations}
  \begin{align}
  \mathbf F^{\mathrm{LL}} &= \mathbf h^{\mathrm{LL}}+\mathbf G^{\mathrm{LL}}+\tfrac12\mathbf Q^{\mathrm{LL}}-\mathbf P^{\mathrm{LL}}\mathbf D^{\mathrm{LL}}\mathbf P^{\mathrm{LL}},\\
  \mathbf F^{\mathrm{LS}} &= \mathbf h^{\mathrm{LS}}+\mathbf G^{\mathrm{LS}}-\mathbf P^{\mathrm{LL}}\mathbf D^{\mathrm{LS}}\mathbf P^{\mathrm{SS}}, \\
  \mathbf F^{\mathrm{SL}}&=(\mathbf F^{\mathrm{LS}})\dag, \\
  \mathbf F^{\mathrm{SS}} &= \mathbf h^{\mathrm{SS}}+\mathbf G^{\mathrm{SS}}+\tfrac12\mathbf Q^{\mathrm{SS}}-\mathbf P^{\mathrm{SS}}\mathbf D^{\mathrm{SS}}\mathbf P^{\mathrm{SS}}.
  \end{align}
  \label{eq:fock}
\end{subequations}
where 
$\mathbf G=\mathbf J-\mathbf K$ is the Dirac--Coulomb
two-electron matrix,
\begin{equation}
  J^{XX}_{\mu\nu} = \sum_{Z\in\{\mathrm L,\mathrm S\}}\sum_{\kappa\tau}(\mu^X\nu^X|\kappa^Z\tau^Z)\,D^{ZZ}_{\tau\kappa},
  \qquad
  K^{XY}_{\mu\nu} = \sum_{\kappa\tau}(\mu^X\tau^X|\kappa^Y\nu^Y)\,D^{XY}_{\tau\kappa},
\end{equation}
and the elements of the four blocks are
\begin{subequations}
\label{eq:fockao}
\begin{align}
  F^{\mathrm{LL}}_{\mu\nu} &= V_{\mu\nu}
   + \sum_{\kappa\tau}D^{\mathrm{LL}}_{\tau\kappa}
     \bigl[(\mu^{\mathrm L}\nu^{\mathrm L}|\kappa^{\mathrm L}\tau^{\mathrm L})
          -(\mu^{\mathrm L}\tau^{\mathrm L}|\kappa^{\mathrm L}\nu^{\mathrm L})\bigr]
   + \sum_{\kappa\tau}D^{\mathrm{SS}}_{\tau\kappa}
     (\mu^{\mathrm L}\nu^{\mathrm L}|\kappa^{\mathrm S}\tau^{\mathrm S})
   \nonumber\\&\quad
   + \tfrac12 Q^{\mathrm{LL}}_{\mu\nu}
   - \sum_{\kappa\tau}P^{\mathrm{LL}}_{\mu\tau}D^{\mathrm{LL}}_{\tau\kappa}P^{\mathrm{LL}}_{\kappa\nu},
  \label{eq:fockao-ll}\\
  F^{\mathrm{LS}}_{\mu\nu} &= T_{\mu\nu}
   - \sum_{\kappa\tau}D^{\mathrm{LS}}_{\tau\kappa}
     (\mu^{\mathrm L}\tau^{\mathrm L}|\kappa^{\mathrm S}\nu^{\mathrm S})
   - \sum_{\kappa\tau}P^{\mathrm{LL}}_{\mu\tau}D^{\mathrm{LS}}_{\tau\kappa}P^{\mathrm{SS}}_{\kappa\nu},
  \label{eq:fockao-ls}\\
  F^{\mathrm{SL}}_{\mu\nu} &= T_{\mu\nu}
   - \sum_{\kappa\tau}D^{\mathrm{SL}}_{\tau\kappa}
     (\mu^{\mathrm S}\tau^{\mathrm S}|\kappa^{\mathrm L}\nu^{\mathrm L})
   - \sum_{\kappa\tau}P^{\mathrm{SS}}_{\mu\tau}D^{\mathrm{SL}}_{\tau\kappa}P^{\mathrm{LL}}_{\kappa\nu},
  \label{eq:fockao-sl}\\
  F^{\mathrm{SS}}_{\mu\nu} &= \tfrac{1}{4c^2}W_{\mu\nu}-T_{\mu\nu}
   + \sum_{\kappa\tau}D^{\mathrm{SS}}_{\tau\kappa}
     \bigl[(\mu^{\mathrm S}\nu^{\mathrm S}|\kappa^{\mathrm S}\tau^{\mathrm S})
          -(\mu^{\mathrm S}\tau^{\mathrm S}|\kappa^{\mathrm S}\nu^{\mathrm S})\bigr]
   + \sum_{\kappa\tau}D^{\mathrm{LL}}_{\tau\kappa}
     (\mu^{\mathrm S}\nu^{\mathrm S}|\kappa^{\mathrm L}\tau^{\mathrm L})
   \nonumber\\&\quad
   + \tfrac12 Q^{\mathrm{SS}}_{\mu\nu}
   - \sum_{\kappa\tau}P^{\mathrm{SS}}_{\mu\tau}D^{\mathrm{SS}}_{\tau\kappa}P^{\mathrm{SS}}_{\kappa\nu}.
  \label{eq:fockao-ss}
\end{align}
\end{subequations}
where $\mathbf P$ and $\mathbf Q$ are the DSE (dipole and quadrupole) matrices and are block diagonal,
\begin{equation}
  \mathbf P = \begin{pmatrix}\mathbf P^{\mathrm{LL}} & \mathbf 0\\ \mathbf 0 & \mathbf P^{\mathrm{SS}}\end{pmatrix},
  \qquad
  \mathbf Q = \begin{pmatrix}\mathbf Q^{\mathrm{LL}} & \mathbf 0\\ \mathbf 0 & \mathbf Q^{\mathrm{SS}}\end{pmatrix},
  \label{eq:pqblock}
\end{equation}
with elements, for $\lam=(\lambda_x,\lambda_y,\lambda_z)$ and
$x_k\in\{x,y,z\}$,
\begin{subequations}
\label{eq:pqao}
\begin{align}
  P^{\mathrm{LL}}_{\mu\nu} &= \sum_{k}\lambda_k\braket{\chi_\mu|x_k|\chi_\nu},
  &
  P^{\mathrm{SS}}_{\mu\nu} &= \frac{1}{4c^2}\sum_{k}\lambda_k
     \braket{\bm\sigma\cdot\hat{\mathbf p}\chi_\mu|x_k|\bm\sigma\cdot\hat{\mathbf p}\chi_\nu},
  \label{eq:pao}\\
  Q^{\mathrm{LL}}_{\mu\nu} &= \sum_{kl}\lambda_k\lambda_l\braket{\chi_\mu|x_kx_l|\chi_\nu},
  &
  Q^{\mathrm{SS}}_{\mu\nu} &= \frac{1}{4c^2}\sum_{kl}\lambda_k\lambda_l
     \braket{\bm\sigma\cdot\hat{\mathbf p}\chi_\mu|x_kx_l|\bm\sigma\cdot\hat{\mathbf p}\chi_\nu}
  \label{eq:qao}\\
  & & &\approx \sum_{\kappa\tau}P^{\mathrm{SS}}_{\mu\kappa}
     \bigl[(\mathbf S^{\mathrm{SS}})^{-1}\bigr]_{\kappa\tau}P^{\mathrm{SS}}_{\tau\nu},
  \qquad \mathbf S^{\mathrm{SS}}=\frac{1}{2c^2}\mathbf T .
  \label{eq:qssri}
\end{align}
\end{subequations}
The terms in the Fock matrix of Eq~\ref{eq:fock} without $\mathbf P$ or $\mathbf Q$ form the cavity-free
Dirac--Coulomb Fock matrix. The terms containing $\mathbf Q$ and $\mathbf P$
are the DSE contributions: (i)~a one-electron
quadrupole term $\tfrac12\mathbf Q$ in the LL and SS blocks and (ii)~a DSE
exchange term $-\mathbf P\mathbf D\mathbf P$ in all four blocks. Because
$\mathbf P$ and $\mathbf Q$ are block diagonal (eq~\ref{eq:pqblock}), the
off-diagonal blocks have no quadrupole term. They nevertheless contain a
DSE exchange term, through the large--small density $\mathbf D^{\mathrm{LS}}$,
which connects $\mathbf P^{\mathrm{LL}}$ and $\mathbf P^{\mathrm{SS}}$.

\subsection{Relativistic Polaritonic Algebraic Diagrammatic Construction Theory}

With canonical 4c-QED-DHF spinors, eq~\ref{eq:hcs} is normal ordered with
respect to $\ket{\Phi}\otimes\ket{0}$ and partitioned as
$\Hop=\Hop_0+\Hop_1$,
\begin{align}
  \Hop_0 &= \sum_p\varepsilon_p\{\cre p\ann p\} + \omega\,\bc\ba, \\
  \Hop_1 &= \tfrac14\sum_{pqrs}\asyl pqrs\{\cre p\cre q\ann s\ann r\}
         - \sq\sum_{pq}\dt_{pq}\,\cre p\ann q\,(\bc+\ba),
  \label{eq:h1}
\end{align}
Both the dressed two-electron
fluctuation and the bilinear coupling are counted as first order.\cite{bauer2023perturbation} The first-order wave function is
\begin{equation}
  \ket{\Psi_0^{(1)}} = -\tfrac14\sum_{ijab} t_{ijab}\cc\,\ket{\Phi_{ij}^{ab}}\otimes\ket 0
     + \sum_{ia}\sq\,\theta_{ia}\cc\,\ket{\Phi_i^a}\otimes\ket 1 .
     \label{eq:psi1}
\end{equation}
with the double-excitation and the photon-dressed single-excitation amplitudes
\begin{equation}
  t_{ijab} = \frac{\asyl ijab}{\varepsilon_a+\varepsilon_b-\varepsilon_i-\varepsilon_j},
  \qquad
  \theta_{ia} = \frac{\dt_{ia}}{\Delta_{ia}+\omega},
  \qquad \Delta_{ia}=\varepsilon_a-\varepsilon_i .
  \label{eq:amps}
\end{equation}
The second component of eq~\ref{eq:psi1}, an electronic single excitation
accompanied by one photon, is specific to cavity QED. Because $\Hop_1$ cannot create a double excitation together with a photon, no
$\ket{\Phi_{ij}^{ab}}\otimes\ket1$ component appears at first order. The
second-order energy is\cite{bauer2023perturbation}
\begin{equation}
  E^{(2)}_{\mathrm{QED\text{-}MP2}} = -\frac14\sum_{ijab}\frac{\lvert\asyl ijab\rvert^2}{\varepsilon_a+\varepsilon_b-\varepsilon_i-\varepsilon_j}
     \;-\; \frac{\omega}{2}\sum_{ia}\frac{\lvert d_{ia}\rvert^2}{\Delta_{ia}+\omega}.
  \label{eq:emp2}
\end{equation}

\begin{figure}
    \centering
    \begin{subfigure}[b]{0.46\textwidth}
    \centering
    \includegraphics[width=\textwidth]{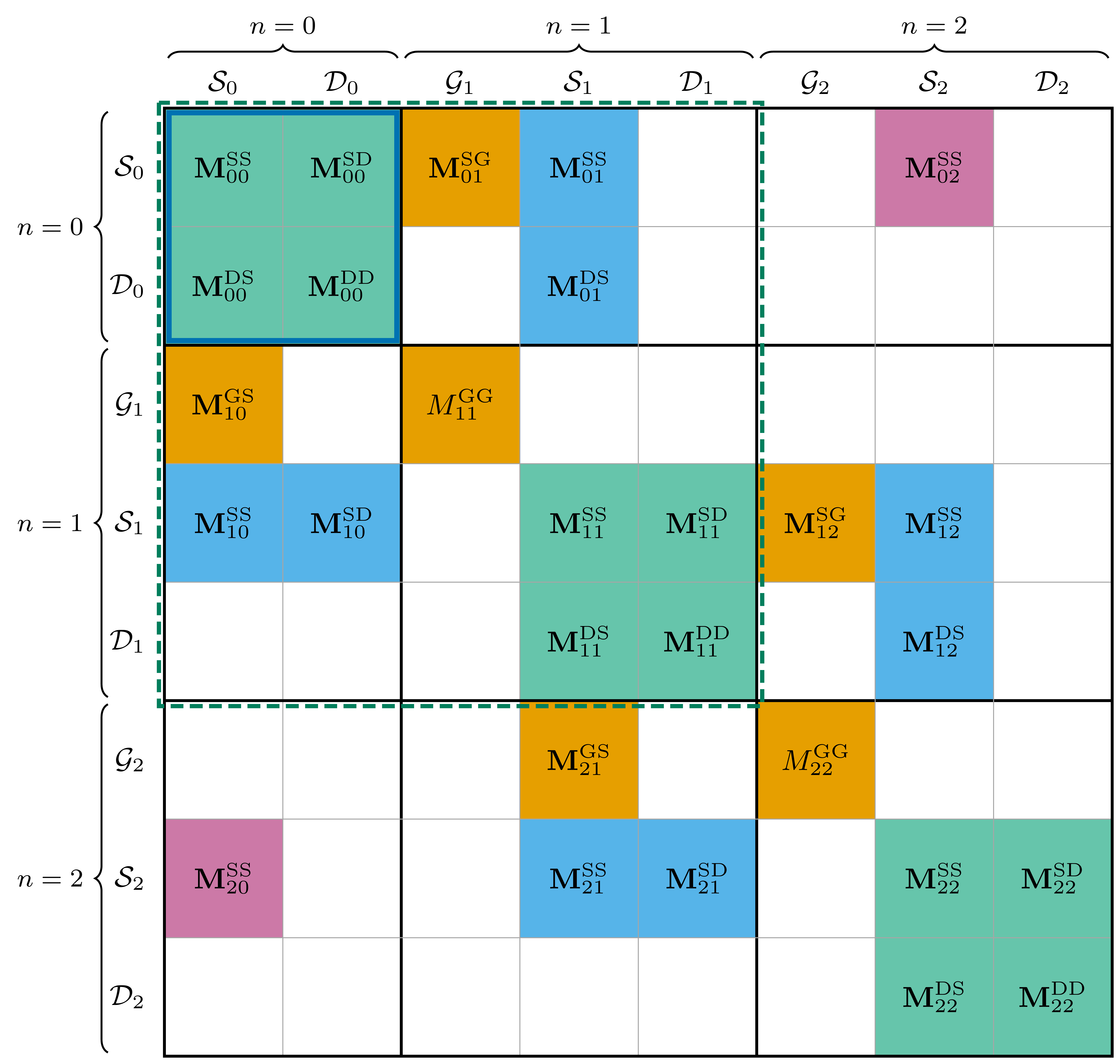}
    \caption{}
    \label{fig:adc_blocks_a}
    \end{subfigure}
    \begin{subfigure}[b]{0.46\textwidth}
    \centering
    \includegraphics[width=\textwidth]{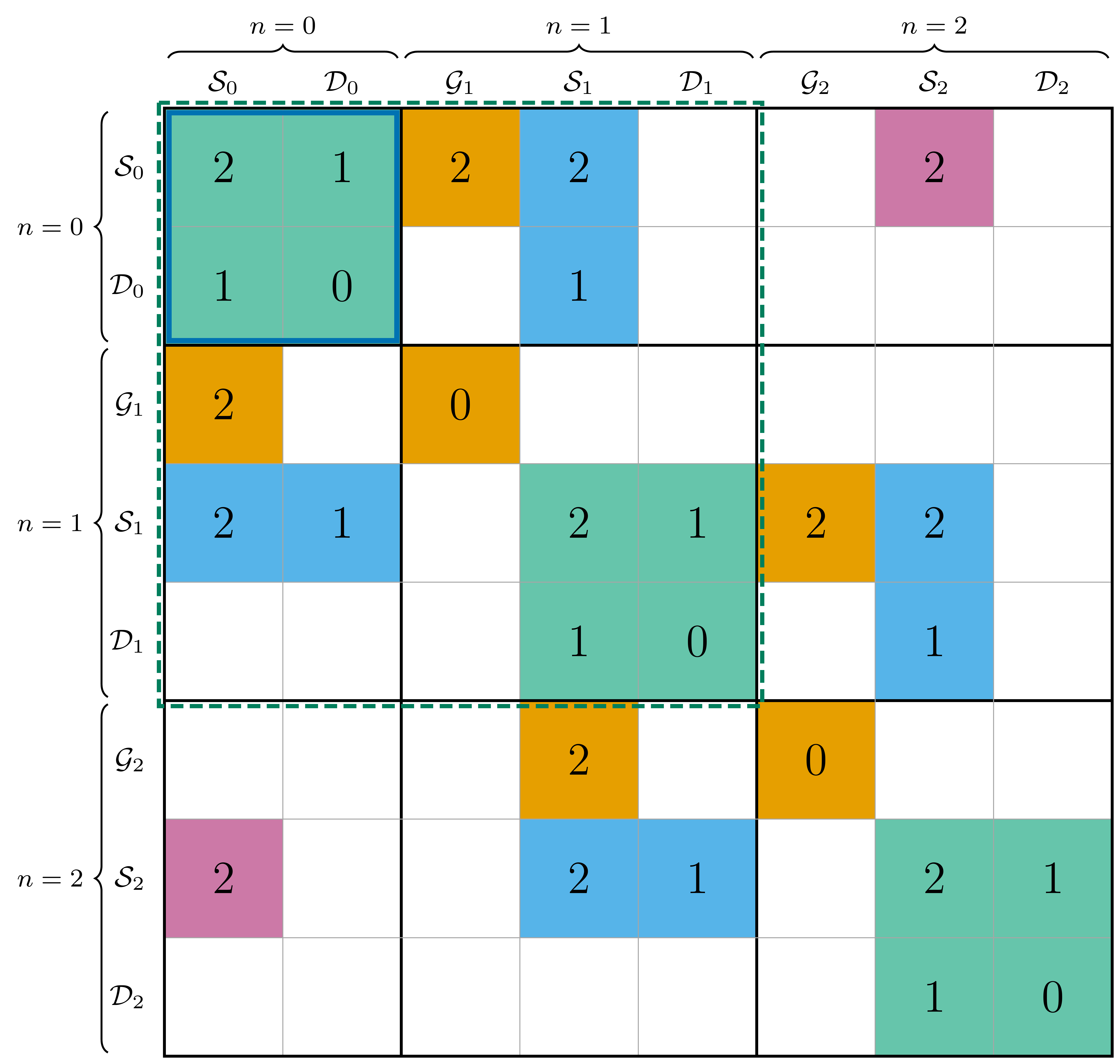}
    \caption{}
    \label{fig:adc_blocks_b}
    \end{subfigure}
    
    \caption{(a) Block structure of the 4c-QED-ADC(2) matrix up to double-photon dispersion and (b) perturbative order retained in each block. The solid frame marks the cavity-free ADC(2) matrix ($n=0$) and the dashed frame the matrix for single-photon dispersion ($n_{\max}=1$).}
    \label{fig:adc_blocks}
\end{figure}

The configuration space is the direct product of the electronic particle--hole (ph) and two-particle--two-hole (2p2h) classes with the
photon-number states $\ket n$, $n=0,1,\dots,n_{\max}$, together with the
photon-dressed ground states. For $n_{\max}=2$ the space consists of the
eight sectors
\begin{equation}
  \underbrace{\mathcal S_0,\ \mathcal D_0}_{n=0},\quad
  \underbrace{\mathcal G_1,\ \mathcal S_1,\ \mathcal D_1}_{n=1},\quad
  \underbrace{\mathcal G_2,\ \mathcal S_2,\ \mathcal D_2}_{n=2},
  \label{eq:sectors}
\end{equation}
where $\mathcal S_n=\{\ket{\tilde\Psi_{ia,n}}\}$ and
$\mathcal D_n=\{\ket{\tilde\Psi_{ijab,n}}\}$ are the ph and
2p2h intermediate states with $n$ photons, and $\mathcal
G_n=\{\ket{\tilde\Psi_{0,n}}\}$ is the electronic ground state with $n\geq1$
photons. The purely electronic ground state $\mathcal G_0$ is excluded,
since it is the reference. The precursor states are generated by applying the electronic excitation operators $\hat C_J\in\{\cre a\ann i,\ \cre a\cre b\ann j\ann i\}$ to the 4c-QED-MP ground state. They are orthogonalized to the ground-state sectors, and the classes are orthonormalized using Gram-Schmidt and a symmetric orthogonalization procedure, as in the cavity-free standard intermediate state representation.\cite{schirmer1982beyond,adc2,chakraborty2026low} The secular matrix
\begin{equation}
  M_{IJ} = \braket{\tilde\Psi_I|\Hop-E_0|\tilde\Psi_J}
\end{equation}
is Hermitian. It is expanded in orders of $\Hop_1$ and truncated according
to the strict ADC(2).\cite{schirmer1982beyond, adc2, bauer2023perturbation}
ph--ph blocks through second order, ph--2p2h couplings through first order, 2p2h--2p2h blocks in zeroth order, and ph--ground-state photon couplings through second order. The eigenvalue problem
\begin{equation}
  \M\mathbf Y = \mathbf Y\bm\Omega,\qquad \mathbf Y\dag\mathbf Y=\mathbf 1,
  \label{eq:secular}
\end{equation}
gives the polaritonic excitation energies $\Omega_n$ relative to the
4c-QED-MP2 ground state, $E_n=E_0+\Omega_n$, and the eigenvectors $\mathbf
Y_n$, partitioned according to Eq~\ref{eq:sectors}. In the complex spinor
basis, $\M$ is a complex Hermitian matrix. The block structure of the 4c-QED-ADC(2) secular matrix is shown in Figure~\ref{fig:adc_blocks}.

Figure~\ref{fig:adc_blocks} shows the structure of the 4c-QED-ADC(2) matrix for
up to two photons. The configuration space is built from three types of
states for every photon number $n$. The photon states $\mathcal G_n$
(the ground state with $n$ photons), the single excitations $\mathcal S_n$
(ph) and the double excitations $\mathcal D_n$ (2p2h). The reference
$\Psi_0$ itself is not part of the space, so there is no $\mathcal G_0$. Each diagonal photon sector (green) is a copy of the electronic ADC(2) matrix, shifted by $n\omega$ and dressed by
the dipole self-energy, and its ph--ph block also contains a second-order light--matter correction that grows with $n$. The $n=0$ block (solid frame) is the cavity-free ADC(2) matrix with its usual perturbative orders\corr{:} second order in the ph--ph block, first order in the ph--2p2h coupling and zeroth order in the 2p2h--2p2h block. Neighboring photon sectors (blue) are linked by the bilinear light--matter coupling, which changes the photon number by one. It connects $\mathcal S_{n+1}$ with $\mathcal S_n$ and with the 2p2h states $\mathcal D_{n}$, which have one photon fewer. The 2p2h states of different photon sectors are not coupled at this order. The orange blocks involve the photon
states $\mathcal G_n$. The coupling between $\mathcal G_1$ and $\mathcal S_0$ is the one that produces the polaritonic splitting when the
cavity frequency is close to an electronic excitation. The purple block
connects $\mathcal S_0$ and $\mathcal S_2$ directly, and it changes the photon number by two and appears only in second order. The elements of the blocks can be represented with the following notations
\begin{equation}
  \begin{gathered}
  (\M^{\mathrm{SS}}_{nm})_{ia,kc}=M_{(ia,n),(kc,m)},\qquad
  (\M^{\mathrm{SD}}_{nm})_{ia,klcd}=M_{(ia,n),(klcd,m)},\qquad
  (\M^{\mathrm{SG}}_{nm})_{ia}=M_{(ia,n),x_m},\\
  (\M^{\mathrm{DS}}_{nm})_{ijab,kc}=M_{(ijab,n),(kc,m)},\qquad
  (\M^{\mathrm{DD}}_{nm})_{ijab,klcd}=M_{(ijab,n),(klcd,m)},\\
  (\M^{\mathrm{GS}}_{nm})_{kc}=M_{x_n,(kc,m)},\qquad
  M^{\mathrm{GG}}_{nm}=M_{x_n,x_m},
  \end{gathered}
  \label{eq:blockmatrix-map}
\end{equation}
where $M_{(ia,n),(jb,m)}=\braket{\tilde\Psi_{ia,n}|\Hop-E_0|\tilde\Psi_{jb,m}}$
and $M_{x_n,(jb,m)}=\braket{\tilde\Psi_{0,n}|\Hop-E_0|\tilde\Psi_{jb,m}}$.
The lowest eigenpairs of eq~\ref{eq:secular} are obtained iteratively using the Davidson procedure, so $\M$ is needed only in the form of matrix-vector products $\bm\sigma=\M\mathbf x$. For $n_{\max}=2$, the trial vector $\mathbf{x}$ is partitioned,
according to Eq.~\eqref{eq:sectors}, into components
$\mathbf{X}_n\in\mathcal{S}_n$ and $\mathbf{Y}_n\in\mathcal{D}_n$
for $n=0,1,2$, together with scalar components
$x_n\in\mathcal{G}_n$ for $n=1,2$.
The elements of $\mathbf{X}_n$ and $\mathbf{Y}_n$ are denoted by
$(\mathbf{X}_n)_{ia}$ and $(\mathbf{Y}_n)_{ijab}$, respectively. The products $\bm\sigma=\M\mathbf x$ are partitioned in the same way 
\begin{subequations}
\label{eq:sigma}
\begin{align}
  \bm\sigma_{\mathcal S_0} &= \M^{\mathrm{SS}}_{00}\mathbf X_0 + \M^{\mathrm{SD}}_{00}\mathbf Y_0
    + \M^{\mathrm{SG}}_{01}\,x_1 + \M^{\mathrm{SS}}_{01}\mathbf X_1 + \M^{\mathrm{SS}}_{02}\mathbf X_2,
  \label{eq:sigma-s0}\\
  \bm\sigma_{\mathcal D_0} &= \M^{\mathrm{DS}}_{00}\mathbf X_0 + \M^{\mathrm{DD}}_{00}\mathbf Y_0
    + \M^{\mathrm{DS}}_{01}\mathbf X_1,
  \label{eq:sigma-d0}\\
  \sigma_{\mathcal G_1} &= \M^{\mathrm{GS}}_{10}\mathbf X_0 + M^{\mathrm{GG}}_{11}\,x_1,
  \label{eq:sigma-g1}\\
  \bm\sigma_{\mathcal S_1} &= \M^{\mathrm{SS}}_{10}\mathbf X_0 + \M^{\mathrm{SD}}_{10}\mathbf Y_0
    + \M^{\mathrm{SS}}_{11}\mathbf X_1 + \M^{\mathrm{SD}}_{11}\mathbf Y_1
    + \M^{\mathrm{SG}}_{12}\,x_2 + \M^{\mathrm{SS}}_{12}\mathbf X_2,
  \label{eq:sigma-s1}\\
  \bm\sigma_{\mathcal D_1} &= \M^{\mathrm{DS}}_{11}\mathbf X_1 + \M^{\mathrm{DD}}_{11}\mathbf Y_1
    + \M^{\mathrm{DS}}_{12}\mathbf X_2,
  \label{eq:sigma-d1}\\
  \sigma_{\mathcal G_2} &= \M^{\mathrm{GS}}_{21}\mathbf X_1 + M^{\mathrm{GG}}_{22}\,x_2,
  \label{eq:sigma-g2s}\\
  \bm\sigma_{\mathcal S_2} &= \M^{\mathrm{SS}}_{20}\mathbf X_0 + \M^{\mathrm{SS}}_{21}\mathbf X_1
    + \M^{\mathrm{SD}}_{21}\mathbf Y_1 + \M^{\mathrm{SS}}_{22}\mathbf X_2
    + \M^{\mathrm{SD}}_{22}\mathbf Y_2,
  \label{eq:sigma-s2}\\
  \bm\sigma_{\mathcal D_2} &= \M^{\mathrm{DS}}_{22}\mathbf X_2 + \M^{\mathrm{DD}}_{22}\mathbf Y_2 .
  \label{eq:sigma-d2}
\end{align}
\end{subequations}
The programmable expressions for all terms of Eq.~\ref{eq:sigma} are given in the Supporting Information.

\section{Computational Details}
The 4c-QED-ADC(2) has been implemented in our in-house software package BAGH.\cite{dutta2025bagh} All the calculations presented here have been performed using the PySCF\cite{sun2015libcint,sun2018pyscf,sun2020recent,thiam2025comprehensive} interface of BAGH.  All calculations use the four-component Dirac--Coulomb Hamiltonian, with the (SS$|$SS) integrals included, and the Gaunt and Breit terms neglected. Unless stated otherwise, the 4c-QED-ADC(2) calculations include photon states with up to two photons. The cavity is described by a single mode with frequency $\omega_c$ and coupling strength $\lambda$. For the Hg atom, the Dyall.v2z basis set was used. The 46 core electrons were frozen, and only virtual spinors with energies up to 9~a.u.\ were correlated. This gives 34 occupied and 84 virtual spinors in the correlation treatment, and the cavity is polarized along $z$. For AuH, the dyall.v2z basis set was used for Au and the uncontracted cc-pVDZ basis set for H, at a bond length of 1.5238~\AA. Frozen core is applied, and virtual spinors with energies up to 2.0~a.u.\ were correlated, giving 34 occupied and 48 virtual spinors. For selenophene, the dyall.v2z basis set was used for Se, and the uncontracted aug-cc-pVDZ basis set for C and H. The 26 core electrons were frozen, and virtual spinors with energies up to 2~a.u.\ were correlated, giving 36 occupied and 206 virtual spinors. The cavity was polarized along $z$, with $\lambda = 0.01$~a.u.
For the HX (X = Br, I, At) series in Section~\ref{sec:hx_series}, we have used the uncontracted cc-pVDZ basis set for hydrogen and the dyall.v2z basis set for the halogen atom, with bond
lengths of 1.414, 1.609, and 1.72~\AA{} for HBr, HI, and HAt, respectively, and the molecular axis along \(z\). The frozen-core approximation is applied, and 18 valence electrons are correlated in all cases. The virtual space is truncated with an energy cutoff of 30~a.u., which leaves 66, 68, and 96 correlated virtual spinors for HBr, HI, and HAt, respectively.
For the potential energy surface of HBr, the dyall.v2z basis set was used for both atoms. The 28 core electrons of Br were frozen, and the full virtual space was correlated, giving 8 occupied and 148 virtual spinors.

\section{Results and Discussion}
\subsection{Hg atom inside cavity}
We begin by considering an atomic system, which provides a particularly transparent setting for elucidating fundamental relativistic effects such as scalar relativistic corrections and spin-orbit coupling (SOC). To this end, we focus on the mercury (Hg) atom, which has also been employed as a benchmark system in the relativistic QED-DFT implementation of Konecny \textit{et al}.\cite{Konecnyrelativistic2025}  Mercury is an especially well-suited prototype due to its strong SOC, which renders the nominally spin-forbidden $^1S_0 \rightarrow {}^3P_1$ transition weakly allowed, while the corresponding $^3P_0$ and $^3P_2$ transitions remain strictly forbidden by dipole selection rules. 
\begin{figure}[h!]
    \centering

    \begin{subfigure}[b]{0.48\textwidth}
        \centering
        \includegraphics[width=\textwidth]{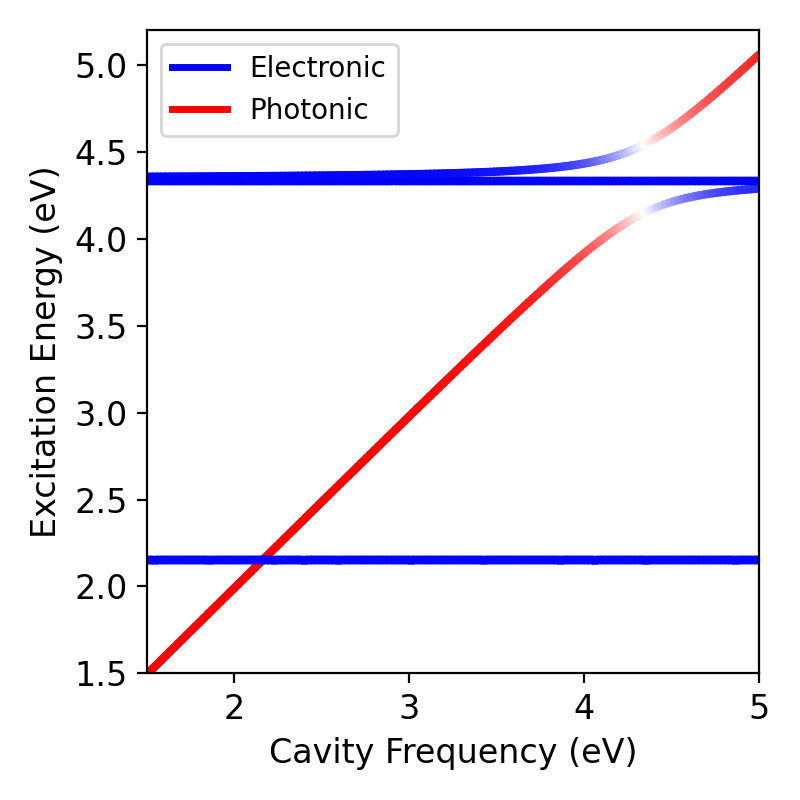}
        \caption{ADC(1), 1c}
        \label{fig:adc1_1c}
    \end{subfigure}
    \hfill
    \begin{subfigure}[b]{0.48\textwidth}
        \centering
        \includegraphics[width=\textwidth]{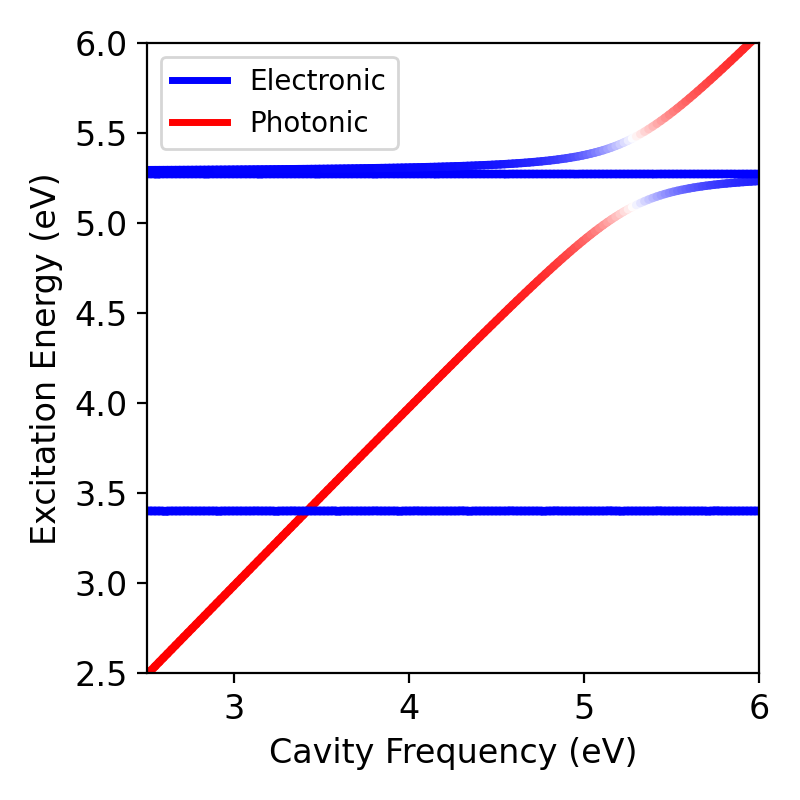}
        \caption{ADC(2), 1c}
        \label{fig:adc2_1c}
    \end{subfigure}

    \vspace{0.8cm}  

    \begin{subfigure}[b]{0.48\textwidth}
        \centering
        \includegraphics[width=\textwidth]{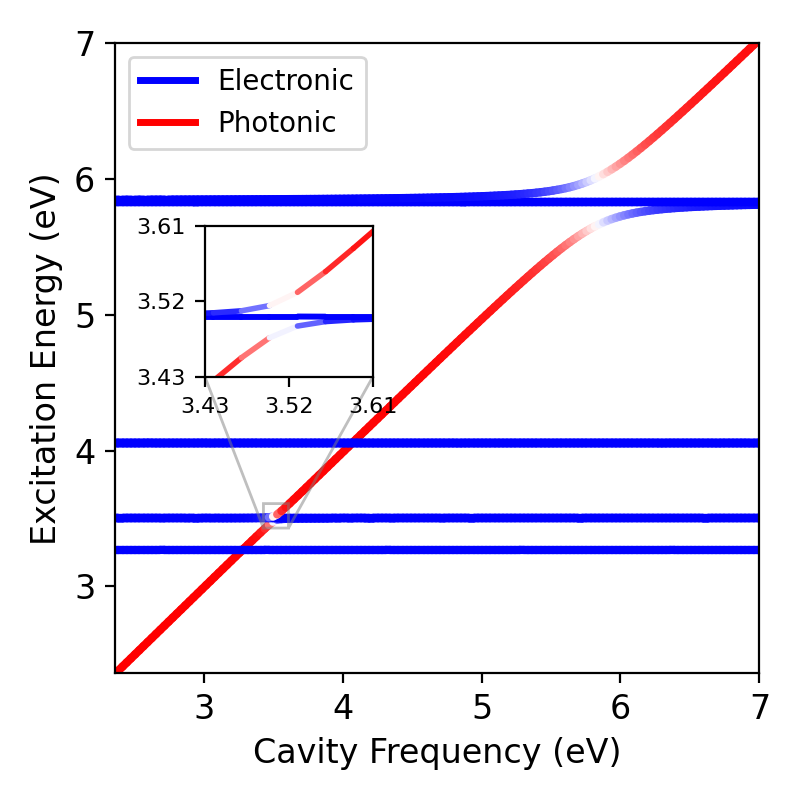}
        \caption{ADC(1), 4c}
        \label{fig:adc1_4c}
    \end{subfigure}
    \hfill
    \begin{subfigure}[b]{0.48\textwidth}
        \centering
        \includegraphics[width=\textwidth]{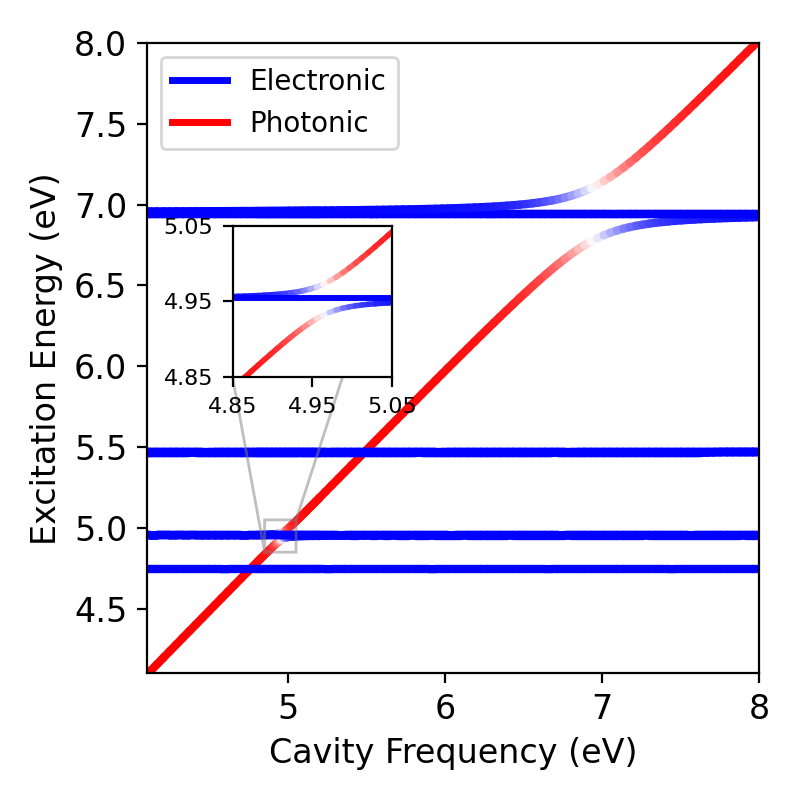}
        \caption{ADC(2), 4c}
        \label{fig:adc2_4c}
    \end{subfigure}

    \caption{The dispersion spectrum of the Hg atom inside a cavity polarized along the $z$-direction with a coupling strength of 0.01 a.u.}
    \label{fig:hg_0.01_dispersion}
\end{figure}
This pronounced SOC-induced state mixing makes Hg an ideal test system for assessing relativistic light–matter interaction effects within our framework. Here, we focus on two specific electronic transitions: the singlet–triplet (S–T) transition $^1S_0 \rightarrow {}^3P_1$ and the singlet–singlet (S–S) transition $^1S_0 \rightarrow {}^1P_1$. Using 4c-EE-ADC(2) calculations in the absence of the cavity, we obtain excitation energies of 4.950 eV and 6.937 eV for the S–T and S–S transitions, respectively. The corresponding experimental values are 4.89 eV and 6.70 eV. The remaining differences between theory and experiment are most likely due to basis-set truncation and missing higher order correlation effects. At this stage, our primary goal is to demonstrate the qualitative behavior of the current implementation. The excited $^3P_J$ and $^1P_J$ states exhibit a $(2J+1)$-fold degeneracy, and consequently, both the $^3P_1$ and $^1P_1$ states are triply degenerate. To examine how strong light–matter coupling modifies these atomic excitations, we employ 4c-QED-EE-ADC(2) for a single Hg atom interacting with a single cavity mode. The cavity frequency is systematically varied across the energy range of the singlet–triplet and singlet–singlet transitions, allowing us to track the evolution of the dispersion spectrum under light-matter coupling conditions. Figure~\ref{fig:hg_0.01_dispersion} shows the dispersion spectra of the Hg atom placed inside an optical cavity with a light–matter coupling strength of 0.01 a.u., where the cavity mode is polarized along the $z$ direction, and also compares the dispersion spectra of the Hg atom obtained using non-relativistic (1c) and fully relativistic (4c) Hamiltonians, evaluated at the QED-ADC(1) and QED-ADC(2) levels of theory. The spectra in Figure~\ref{fig:hg_0.01_dispersion} display the excitation energies as a function of the cavity frequency, with the color scale indicating the relative photonic and electronic character of each state. The lower panel of Figure~\ref{fig:hg_0.01_dispersion} shows the dispersion obtained from fully relativistic (4c) calculations at the ADC(1) and ADC(2) levels. Two distinct polaritonic branches are clearly visible, one forming near the singlet–triplet (S–T) transition and the other around the singlet–singlet (S–S) transition, indicating strong light–matter hybridization in both spectral regions. As shown in the figure, only a single lower and upper polaritonic branch is formed near the resonance regions associated with the S–T and S–S transitions, even though the $^3P_1$ and $^1P_1$ states are triply degenerate. This behavior arises from the directional nature of the transition dipole moments. For both $^3P_1$ and $^1P_1$, only one of the three degenerate components has a transition dipole moment aligned with the polarization direction of the cavity mode and therefore couples strongly to the cavity field. The remaining two degenerate components have dipole moments oriented perpendicularly to the cavity polarization and consequently do not interact with the cavity, remaining essentially unchanged. In contrast, when a non-relativistic Hamiltonian is employed, a different behavior is observed, and only a single polaritonic branch appears, which occurs exclusively near the singlet–singlet (S–S) transition. Beyond these qualitative differences, the non-relativistic treatment also leads to noticeable quantitative deviations. In particular, the S–T and S–S excitation energies in the non-relativistic QED-ADC(2) calculation are red-shifted by 1.553 eV and 1.670 eV, respectively, relative to the fully relativistic results at the QED-ADC(2) level. At the QED-ADC(1) level, a similar trend is observed, where switching from the fully relativistic 4c Hamiltonian to the non-relativistic description leads to a noticeable red-shift of the excitation energies by 1.351 eV for the S–T transition and 1.495 eV for the S–S transition. Another noticeable difference appears when comparing QED-ADC(1) and QED-ADC(2) within the 4c relativistic framework. The S–T and S–S excitation energies differ by 1.447 eV and 1.109 eV, respectively, between the two levels of theory. As a result, the positions of the corresponding polaritonic branches are shifted, highlighting the important role of electron correlation in obtaining accurate excitation energies. It is important to note that the hybrid lower and upper polaritonic states near the S–T and S–S excitation energies do not exhibit equal light–matter contributions exactly at the free-atom excitation energies. Instead, the point at which the photonic and electronic characters become comparable is shifted slightly to higher energies. At the 4c-QED-ADC(2) level, this occurs at approximately 4.963 eV for the S–T transition and about 6.96 eV for the S–S transition.
\begin{figure}[h!]
    \centering
    
    \begin{subfigure}[b]{0.48\textwidth}
        \centering
        \includegraphics[width=\textwidth]{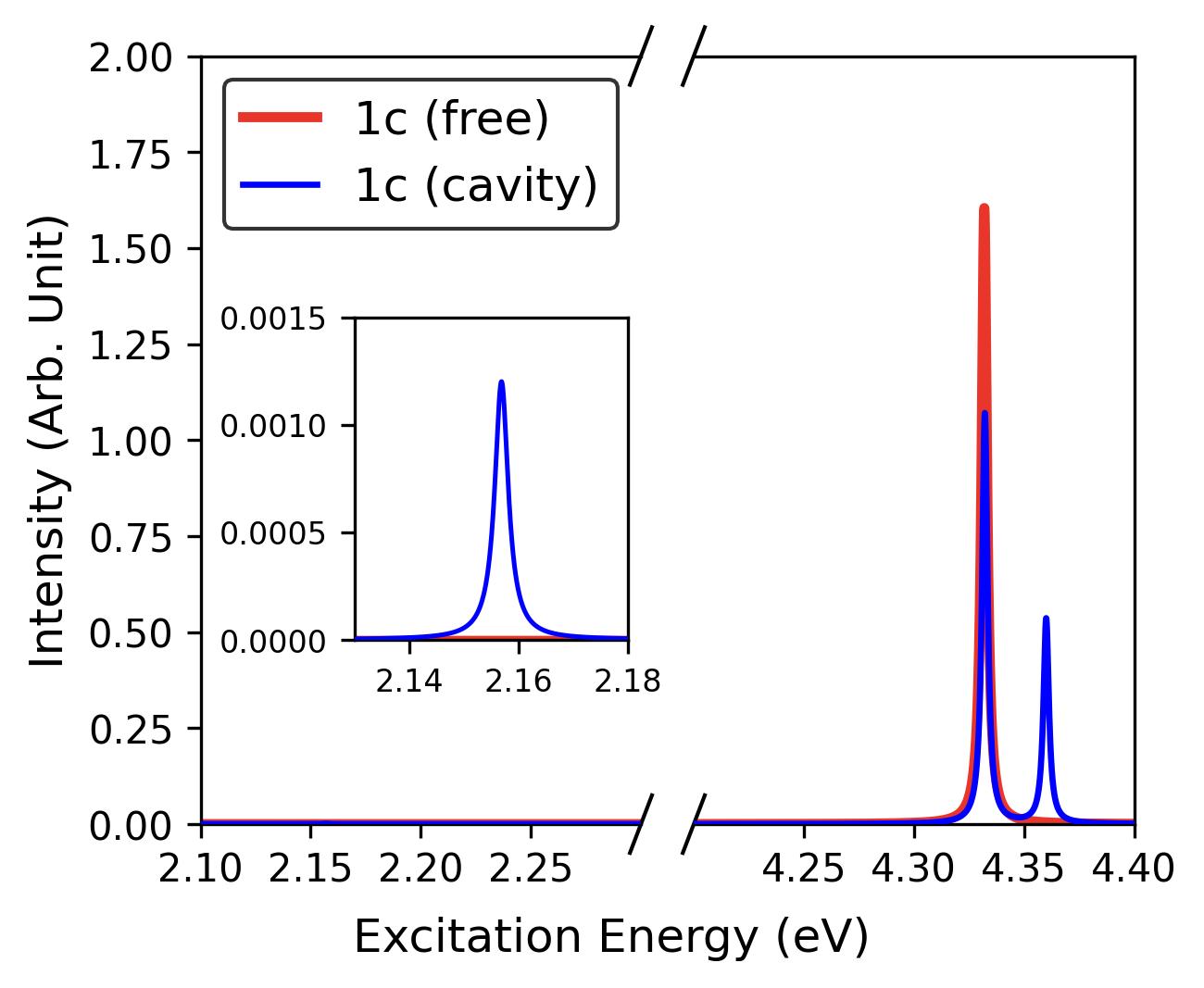}
        \caption{Hg, 1c, ADC(1)}
        \label{fig:hg001_abs_a}
    \end{subfigure}
    \hfill
    \begin{subfigure}[b]{0.48\textwidth}
        \centering
        \includegraphics[width=\textwidth]{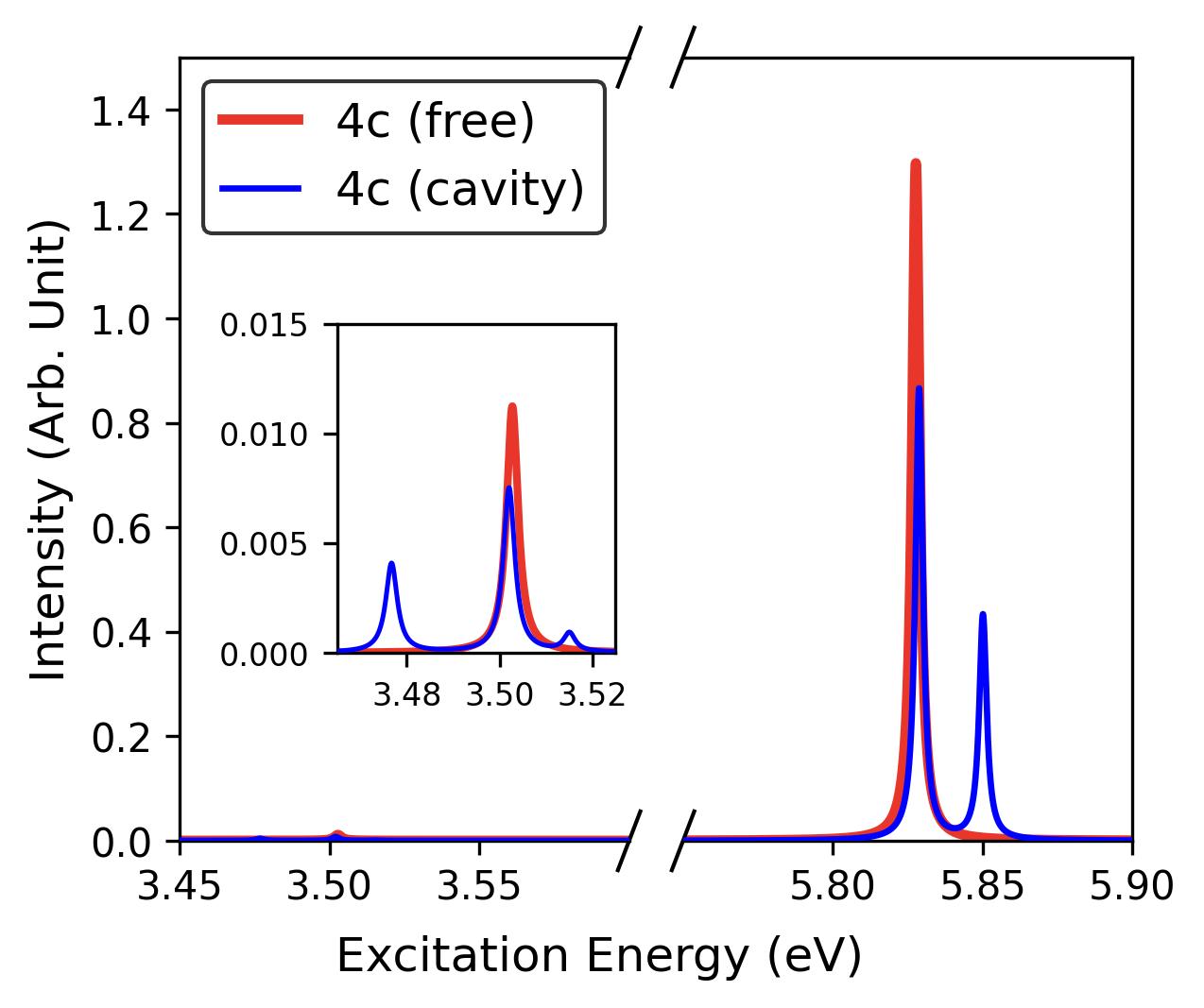}
        \caption{Hg, 4c, ADC(1)}
        \label{fig:hg001_abs_b}
    \end{subfigure}
    
    \begin{subfigure}[b]{0.48\textwidth}
        \centering
        \includegraphics[width=\textwidth]{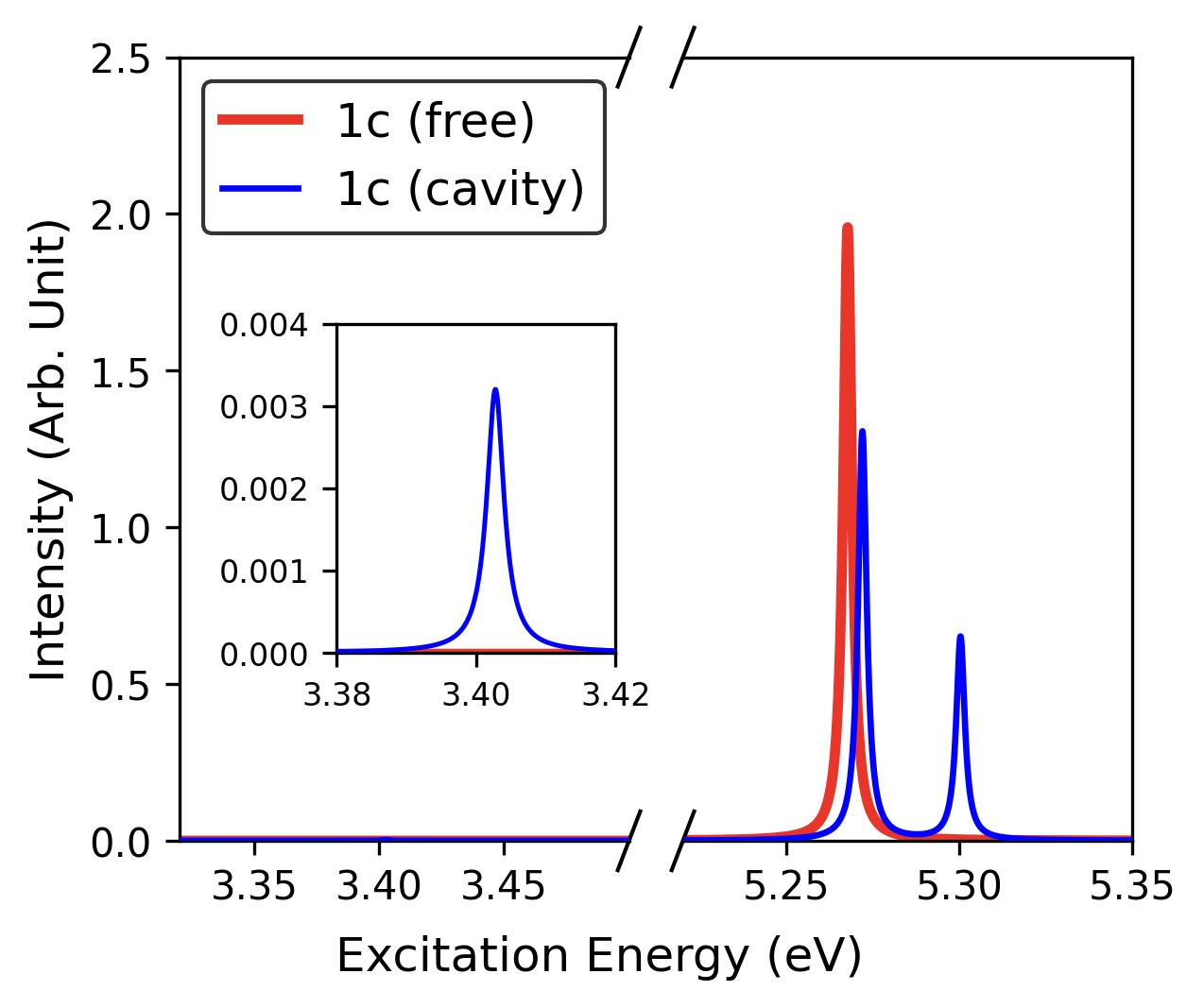}
        \caption{Hg, 1c, ADC(2)}
        \label{fig:hg001_abs_c}
    \end{subfigure}
    \hfill
    \begin{subfigure}[b]{0.48\textwidth}
        \centering
        \includegraphics[width=\textwidth]{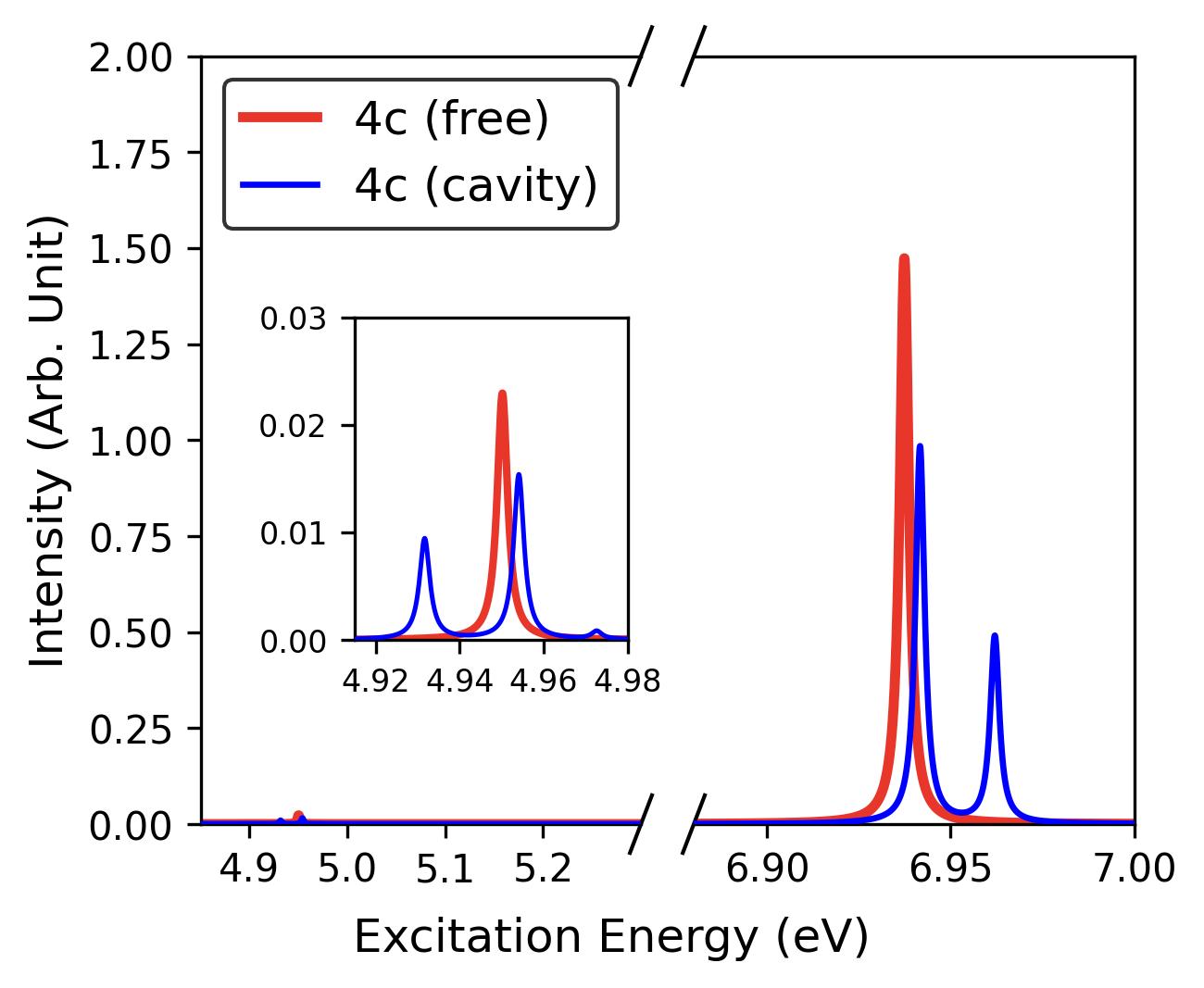}
        \caption{Hg, 4c, ADC(2)}
        \label{fig:hg001_abs_d}
    \end{subfigure}

    \caption{Absorption spectra of Hg atom inside cavity with a coupling strength of 0.01 a.u. and compared with the free spectra.}
    \label{fig:hg_0.01_absorption_4c_1c}
\end{figure}

We have also computed the absorption spectrum of a Hg atom inside an optical cavity, with the cavity frequency tuned to the S–T transition, as shown in Figure~\ref{fig:hg_0.01_absorption_4c_1c}, and the resulting spectrum is compared with that of the corresponding uncoupled (without cavity) system. It can be observed that the absorption spectrum exhibits five distinct peaks when the atom is placed inside the cavity, compared to only two peaks in the uncoupled case, even though the cavity frequency is tuned exclusively to the S–T transition. An examination of the four-component (4c) spectra at the QED-ADC(2) level reveals three distinct peaks in the vicinity of the S–T excitation energy, along with two peaks near the S–S excitation region. The polaritonic peaks in the vicinity of the S–T transition exhibit lower intensities compared to the corresponding free (uncoupled) S–T transition. In addition, one peak arising from two degenerate states with transition dipole moments oriented along mutually perpendicular directions is also observed. This peak possesses an intensity of approximately two-thirds of that of the uncoupled S–T transition. Notably, when the intensities of the LP and UP peaks on either side of the central peak are summed, the total intensity of these three peaks in the vicinity of the S–T transition is slightly larger, by about 10\%, than that of the corresponding uncoupled S-T peak, which shows that the two polaritonic peaks originate from the coupling with the S-T transition; the small excess intensity is borrowed from the S–S transition, as discussed below. It is interesting to note that in the non-relativistic (1c) absorption spectrum, a very low-intensity peak is observed even though the S–T transition is absent in this case. Therefore, this peak is likely due to the mixing of the S–S transition with the cavity mode. The very weak peak near 3.40 eV in the 1c spectrum is predominantly photonic in character, with a contribution exceeding 95\%, and has about one third of the intensity of the LP observed near the S–T region in the 4c spectrum. It should be noted that, in addition to the three peaks observed near the S–T transition in the 4c spectrum, two additional peaks appear in the vicinity of the S–S transition, even though the cavity frequency is far off-resonant with the S–S transition. Among these two peaks, the higher-intensity peak, which closely resembles the uncoupled S–S transition, exhibits approximately two-thirds of the intensity of the uncoupled S–S peak. A similar redistribution of intensity is also observed in the S–T region. However, an additional peak of comparatively lower intensity appears at slightly higher energy. An analysis of its character reveals that it is predominantly electronic in nature, with a very weak but non-zero photonic contribution. Thus, the intensities of the lower and upper polaritons near the S–T transition include a small but noticeable contribution from this shifted S–S transition. 

\begin{figure}[h!]
    \centering
    \begin{subfigure}[b]{0.45\textwidth}
        \centering
        \includegraphics[width=\textwidth]{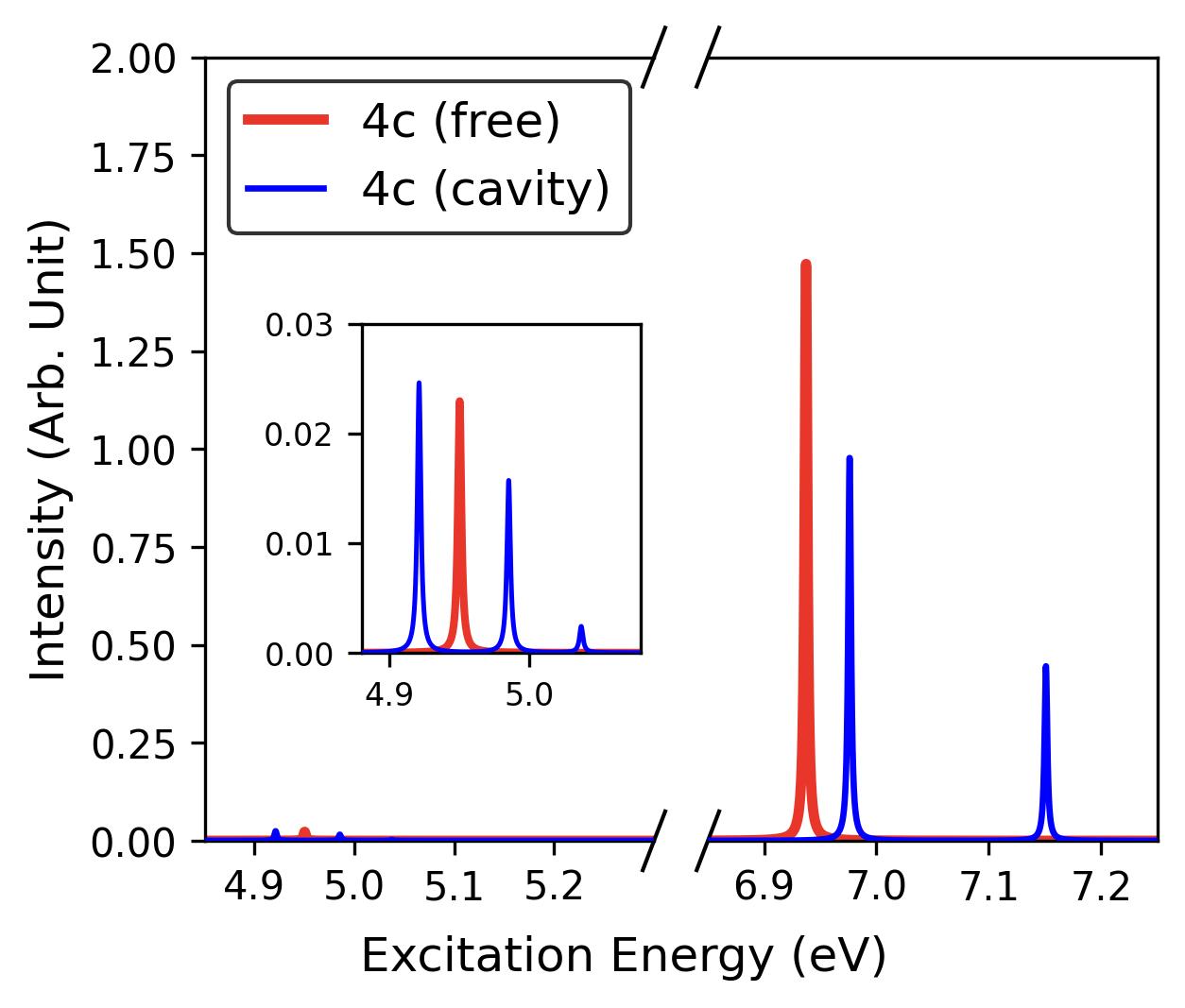}
        \caption{}
        \label{fig:hg_abs_003}
    \end{subfigure}
    \hfill
    \begin{subfigure}[b]{0.45\textwidth}
        \centering
        \includegraphics[width=\textwidth]{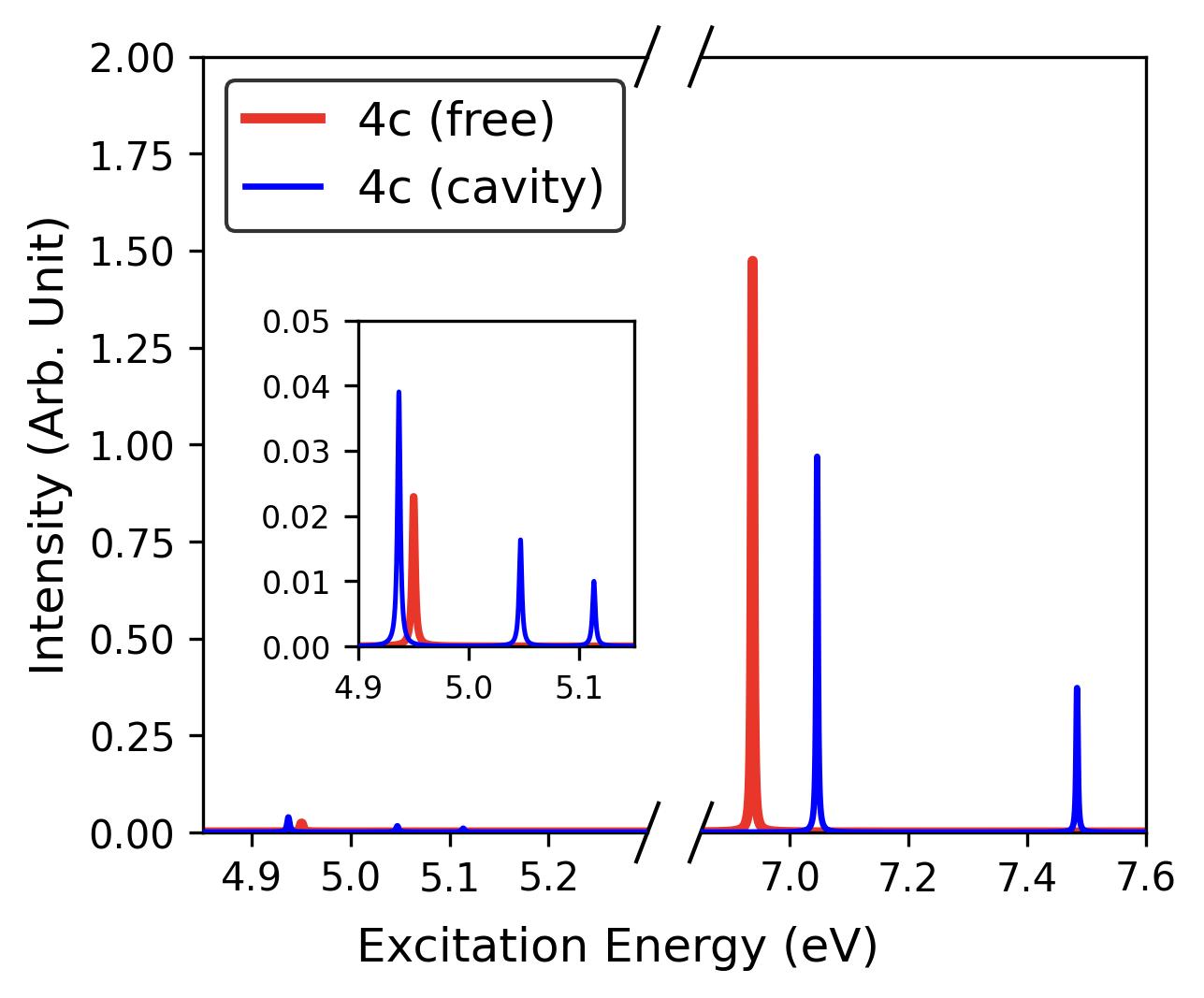}
        \caption{}
        \label{fig:hg_abs_005}
    \end{subfigure}
    
    \caption{Absorption spectra of the Hg atom inside a cavity polarized along the $z$-direction with a coupling strength of (a) 0.03 a.u. and (b) 0.05 a.u., with the cavity frequency set near the S-T transition using 4c-QED-ADC(2).}
    \label{fig:hg_0.03_0.05_absorption}
\end{figure}

Increasing the coupling strength to 0.03 and 0.05 a.u. leads to noticeable modifications in the features of the absorption spectra, including shifts in peak positions and changes in relative intensities. The dispersion spectra of Hg for these coupling strengths are given in Figure S1 of the Supporting Information (SI). The absorption spectra with these coupling strengths are shown in Figure~\ref{fig:hg_0.03_0.05_absorption}. When the coupling strength is increased to 0.03 a.u., an almost analogous behavior is observed compared to the case with a coupling strength of 0.01 a.u. However, it is observed that the LP in the vicinity of the S–T region exhibits a higher intensity than the central peak, which corresponds to two uncoupled states with mutually perpendicular transition dipole moments and it is already slightly more intense than the S–T peak of the free atom. Upon further increasing the coupling strength to 0.05 a.u., the qualitative nature of the absorption spectra changes more significantly, particularly in the S–T region. At a coupling strength of 0.05 a.u., the LP near the S–T transition is about 1.7 times as intense as the S–T peak of the free atom and clearly more intense than the peak associated with the two uncoupled states discussed above. It is noteworthy that, with increasing coupling strength, the point of 50:50 light–matter mixing shifts further toward higher energies (blue shift), and the magnitude of this shift increases with the coupling strength. For instance, the dispersion spectra in Figure S1 of the SI show that equal light–matter mixing occurs at approximately 5.08 eV for 0.03 a.u. and 5.27 eV for 0.05 a.u., compared with 4.963 eV for 0.01 a.u. It is also observed that with an increase in the coupling strength, the separation between the LP and UP increases, and the uncoupled S–T peak, the UP, and the S–S peaks shift to higher energies. Additionally, the separation between the peaks associated with the S–S transition near 7.0 eV also increases. Another important aspect is that, with increasing coupling strength, the shifted low-intensity S–S transition acquires a larger photonic character. As a consequence, the degree of state mixing is enhanced, which in turn leads to an increased intensity of the polaritonic peaks near the S–T transition. This behavior highlights the crucial role of mixing between neighboring electronic states in realistic systems placed inside an optical cavity and also underscores the importance of relativistic effects in accurately describing such light–matter interactions.

    

\subsection{Polaritonic versus non-polaritonic DHF reference}

\begin{figure}[h!]
    \centering

    \begin{subfigure}{0.48\textwidth}
        \centering
        \includegraphics[width=\textwidth]{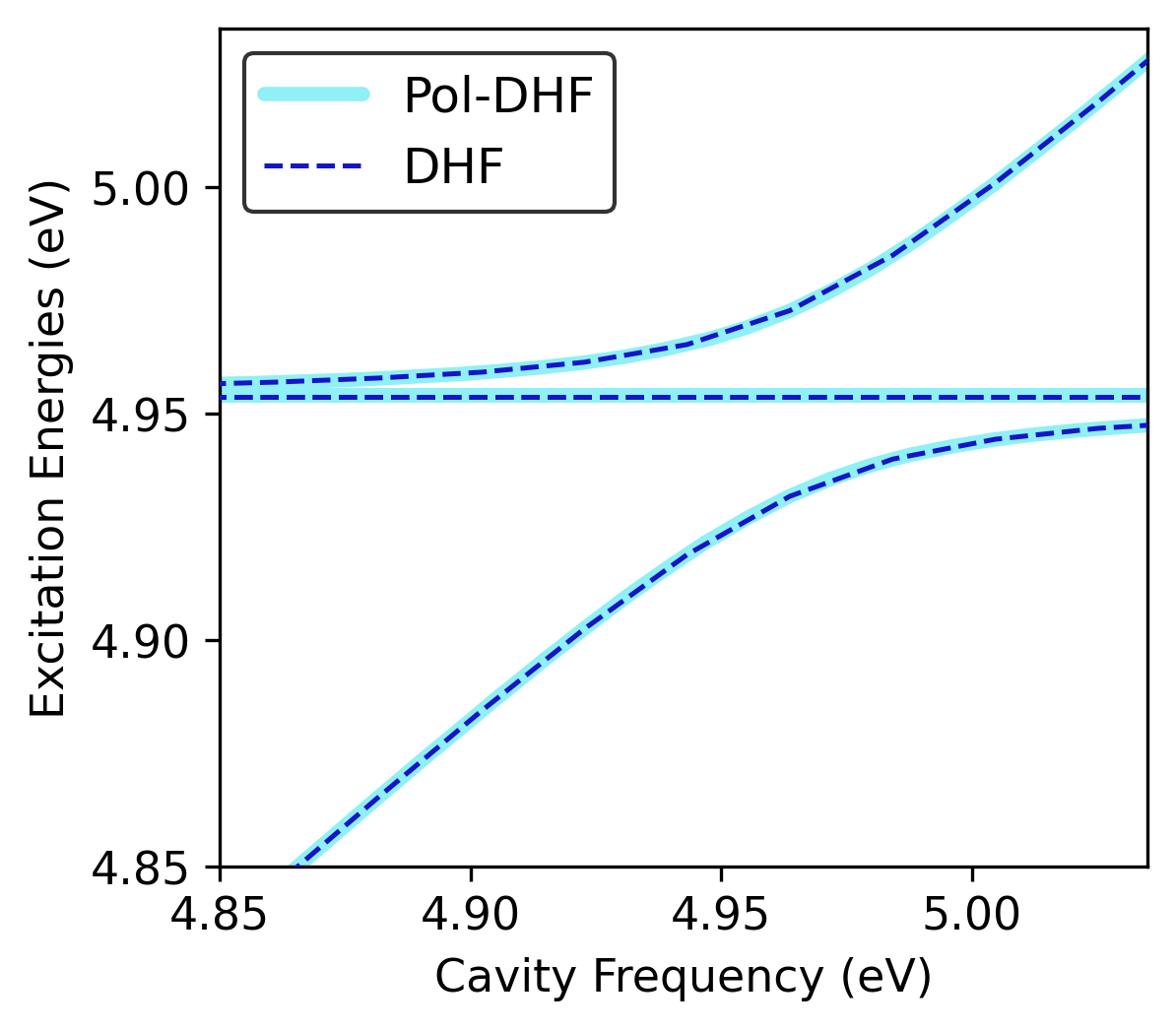}
        \caption{}
        \label{fig:pol_nonpol_001_a}
    \end{subfigure}
    \hfill
    \begin{subfigure}{0.48\textwidth}
        \centering
        \includegraphics[width=\textwidth]{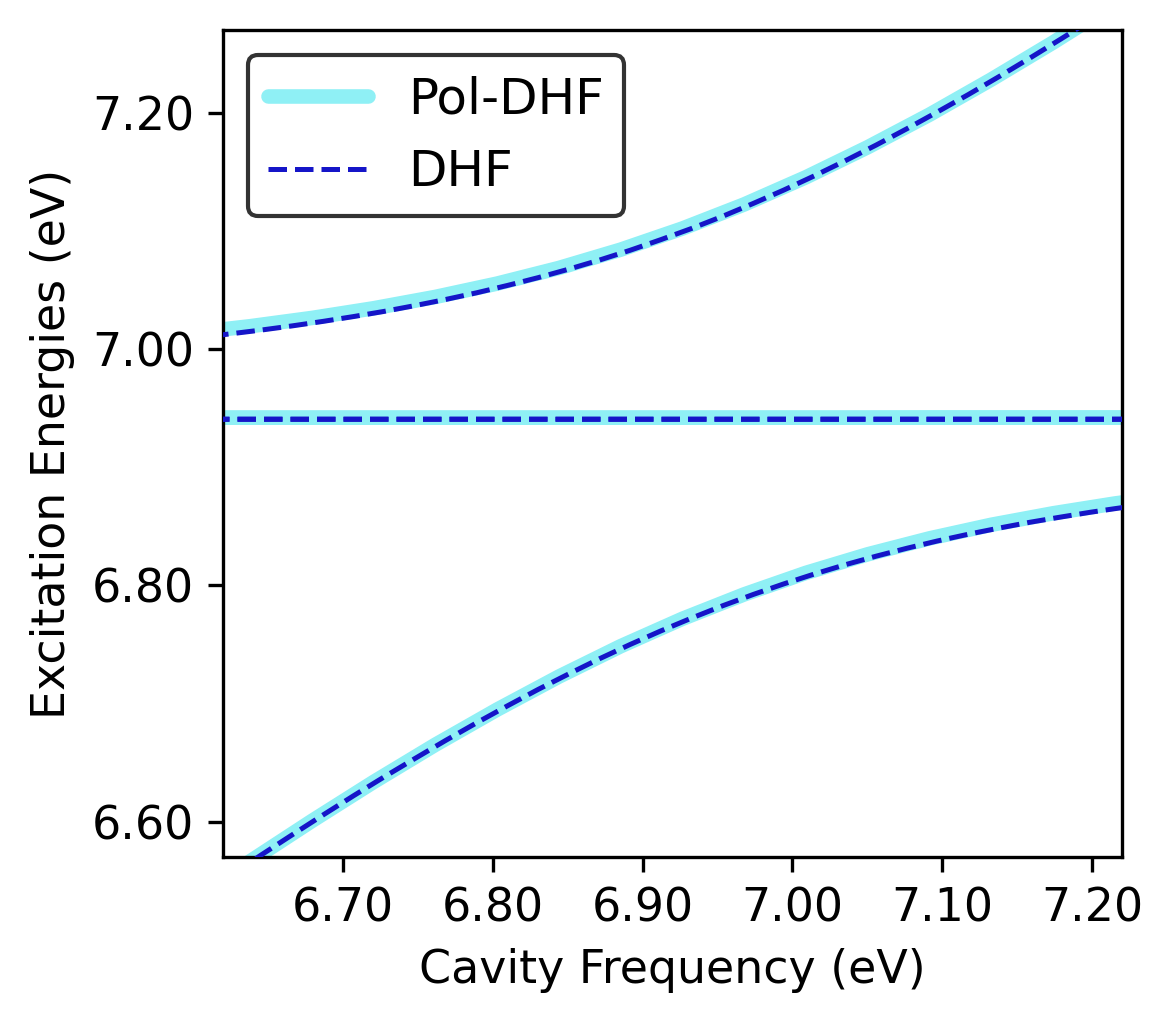} 
        \caption{}
        \label{fig:pol_nonpol_001_b}
    \end{subfigure}

    \caption{Cavity dispersion spectra of Hg near the (a) S–T and (b) S–S transitions at a coupling strength of 0.01 a.u., calculated with the polaritonic 4c-QED-DHF reference (Pol-DHF, solid lines) and the non-polaritonic 4c-DHF reference (DHF, dashed lines).}
    \label{fig:pol_vs_nonpol_0.01}
\end{figure}
In Figure~\ref{fig:pol_vs_nonpol_0.01} we have compared the cavity dispersion spectra of the Hg atom near the S–T and S–S transitions at a light–matter coupling strength of 0.01 a.u., obtained with 4c-QED-ADC(2) on the polaritonic reference (Pol-DHF) and on the non-polaritonic (DHF) reference. Both approaches reproduce the expected Rabi splitting behavior at resonance, with the lower and upper polaritonic branches emerging symmetrically around the bare transition energy. Importantly, the two methods yield nearly indistinguishable spectra in this regime, as indicated by the close overlap of the solid and dashed curves. Near the S–T transition, the Rabi splitting is 40.9~meV with the polaritonic and 41.0~meV with the non-polaritonic reference, and equal light–matter mixing occurs at 4.963 and 4.964~eV, respectively. Near the S–S transition, the splittings are 330.0 and 329.6~meV, and the points of equal mixing differ by only about 5~meV. The components of $^3P_1$ and $^1P_1$ that do not couple to the cavity differ by only 0.3 and 1.3~meV between the two references. This agreement shows that the non-polaritonic DHF reference, in which the cavity enters only at the correlated level through 4c-QED-MP2 and 4c-QED-ADC(2), provides a quantitatively reliable description of the polaritonic excited states. It also shows that, at this coupling strength, including the cavity parameters already in the mean-field reference, changes the polaritons only slightly. As shown for AuH in the next section, the differences between the two references grow with the coupling strength.

\subsection{AuH inside cavity}
\begin{figure*}
    \centering
    \begin{subfigure}[b]{0.9\textwidth}
    \centering
    \includegraphics[width=\textwidth]{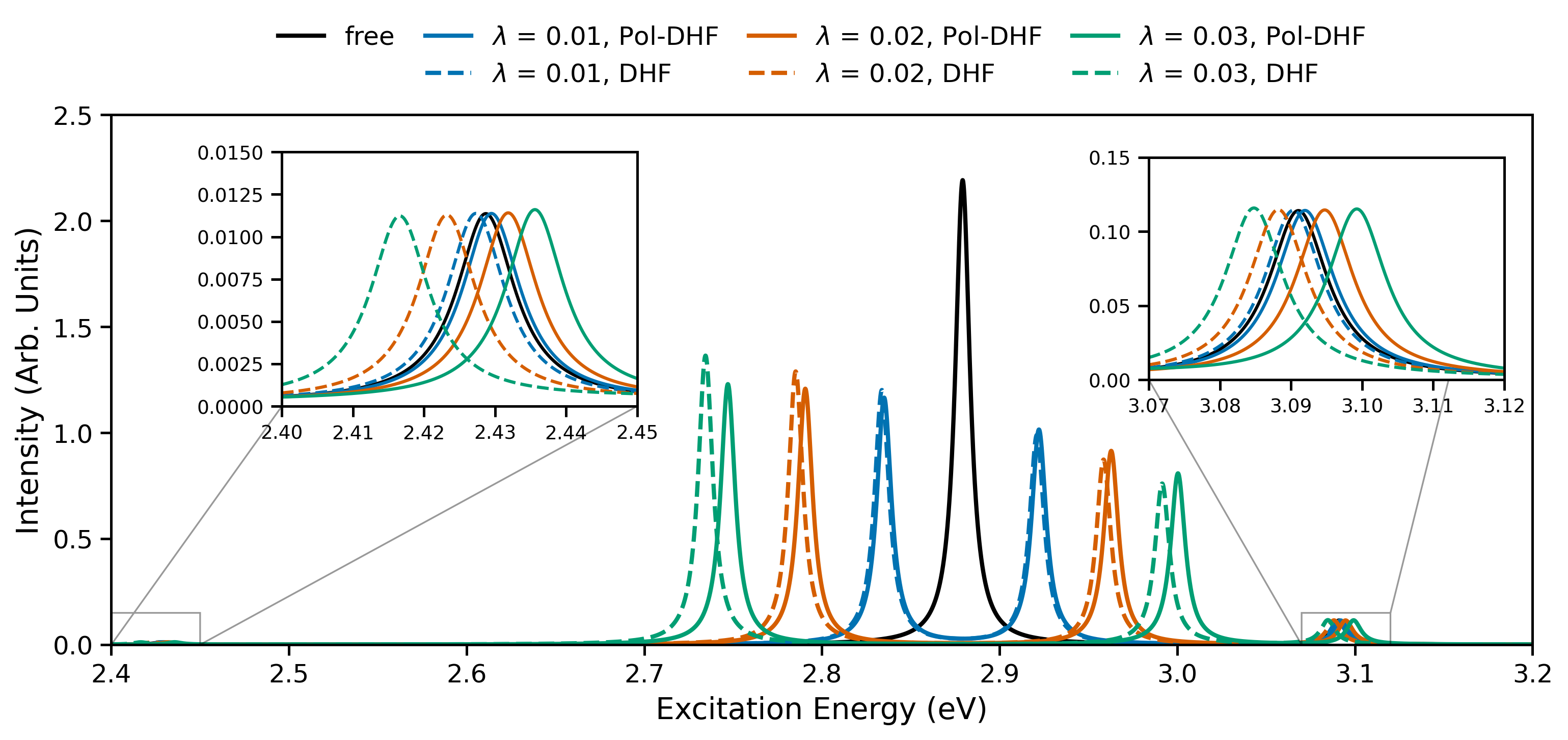}
    \caption{}
    \label{fig:a}
    \end{subfigure}
    
    \caption{The absorption spectra of the AuH inside a cavity with various coupling strengths and polarized along the z-direction, and the cavity frequency is set to resonance with the transition to the $0^{+}$ state. The calculations have been performed using non-polaritonic 4c-DHF and polaritonic 4c-DHF references, denoted as DHF and pol-DHF, respectively, in the figure.}
    \label{fig:auh_0+_coup_variation}
\end{figure*}

Figure~\ref{fig:auh_0+_coup_variation} shows the absorption spectrum of AuH inside a cavity polarized along the bond and tuned to the bright $0^+$ state, for three coupling strengths on the Pol-DHF (solid lines) and DHF (dashed lines) references, together with the spectrum of the free molecule (black line). The strong $0^+$ peak of the free molecule at 2.879~eV splits into two peaks inside the cavity, a lower and an upper polariton. The two peaks move further apart as the coupling increases. The Rabi splitting is 87, 172, and 253~meV on the Pol-DHF reference and 87, 173, and 257~meV on the DHF reference for $\lambda = 0.01$, 0.02, and 0.03~a.u., and it grows almost linearly with $\lambda$. The two references give almost the same splitting, so the strength of this coupling depends only weakly on the reference. The result differs more in the positions of the polaritons depending on the choice of reference. The pair does not remain centered on the free $0^+$ peak but instead shifts to lower energy. The midpoint between the LP and UP lies below the free peak by 0.6, 2.5 and 5.5~meV on the Pol-DHF reference, and by 1.9, 7.3 and 16.2~meV on the DHF reference. This shift grows with $\lambda$, and it is about three times smaller on the Pol-DHF reference. As a result, the solid and dashed lines almost overlap at $\lambda = 0.01$~a.u.\ and separate clearly at $\lambda = 0.03$~a.u. Each polariton peak is much lower in intensity than the free $0^+$ peak, because the intensity of the free state is now shared between the two polaritons. The sharing is not equal, and the LP is always brighter than the UP. The LP is 1.15, 1.3, and 1.5 times brighter than the UP on the Pol-DHF reference, and 1.2, 1.5, and 1.8 times brighter on the DHF reference. The insets of the Figure~\ref{fig:auh_0+_coup_variation} show two weak SOC-induced transitions to the spin-forbidden $\Omega = 1$ state at 2.429~eV and a state at 3.091~eV. Neither peak splits, and their heights remain almost unchanged. This is expected because the transition dipoles of these states lie perpendicular to the bond and therefore perpendicular to the cavity field, so they do not couple directly to the photon. Their positions still change, however, and the shift grows with the coupling. On the Pol-DHF reference, both peaks move to higher energy, by 0.8, 3.2 and 6.9~meV for $1(\mathrm{I})$ and by 0.9, 3.7 and 8.2~meV for the 3.091~eV state. On the DHF reference, both move to lower energy by 1.4, 5.5 and 12.1~meV and by 0.7, 2.8 and 6.3~meV. The cavity therefore changes these states even when no polariton forms. The dipole self-energy raises both the ground and the excited states by large, nearly equal amounts. At $\lambda = 0.03$~a.u.\ on the Pol-DHF reference, they rise by 127 and 134~meV, so the peak moves up by only 6.9~meV. Because the peak shift is a small difference between two large shifts, it is sensitive to how the cavity is treated. On the DHF reference, where the spinors are not relaxed in the cavity, the peak moves down instead.

\begin{figure}[h!]
    \centering

    \begin{subfigure}{0.48\textwidth}
        \centering
        \includegraphics[width=\textwidth]{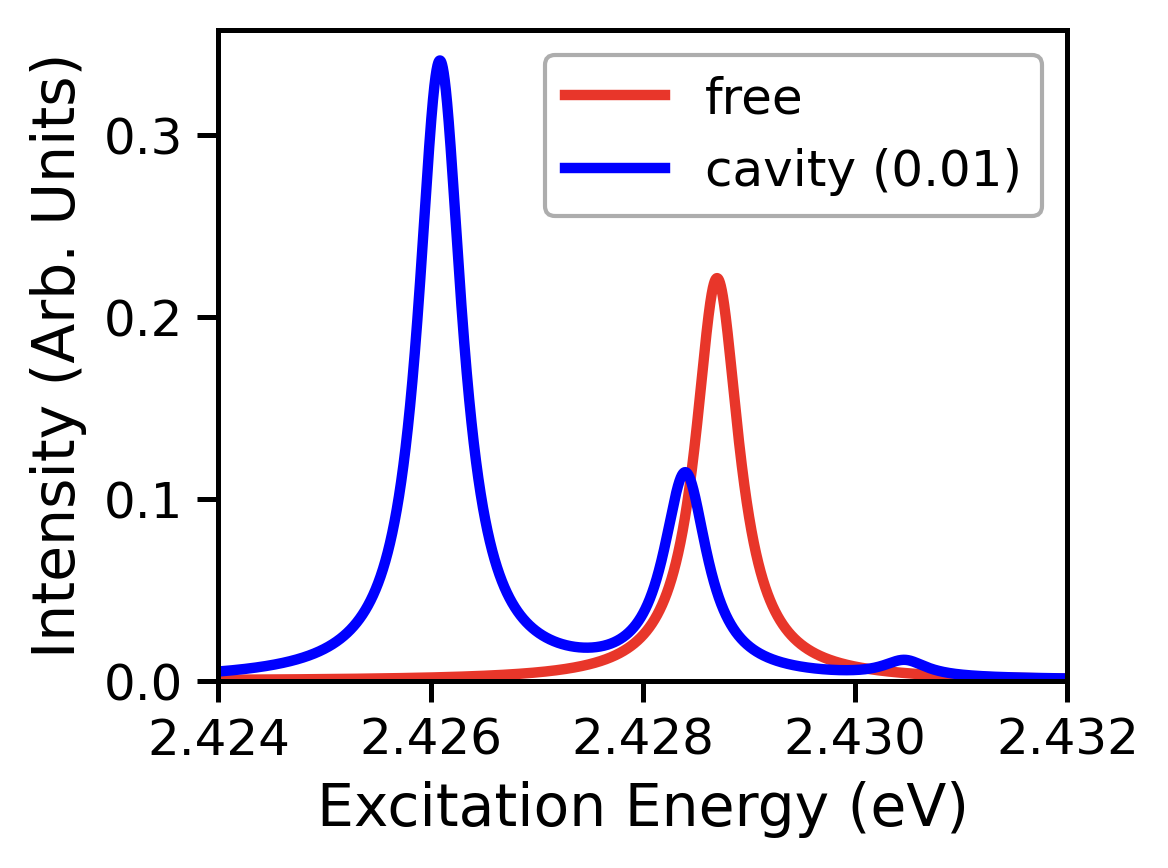}
        \caption{}
        \label{fig:auh_0.01_1Pi_x}
    \end{subfigure}
    \hfill
    \begin{subfigure}{0.48\textwidth}
        \centering
        \includegraphics[width=\textwidth]{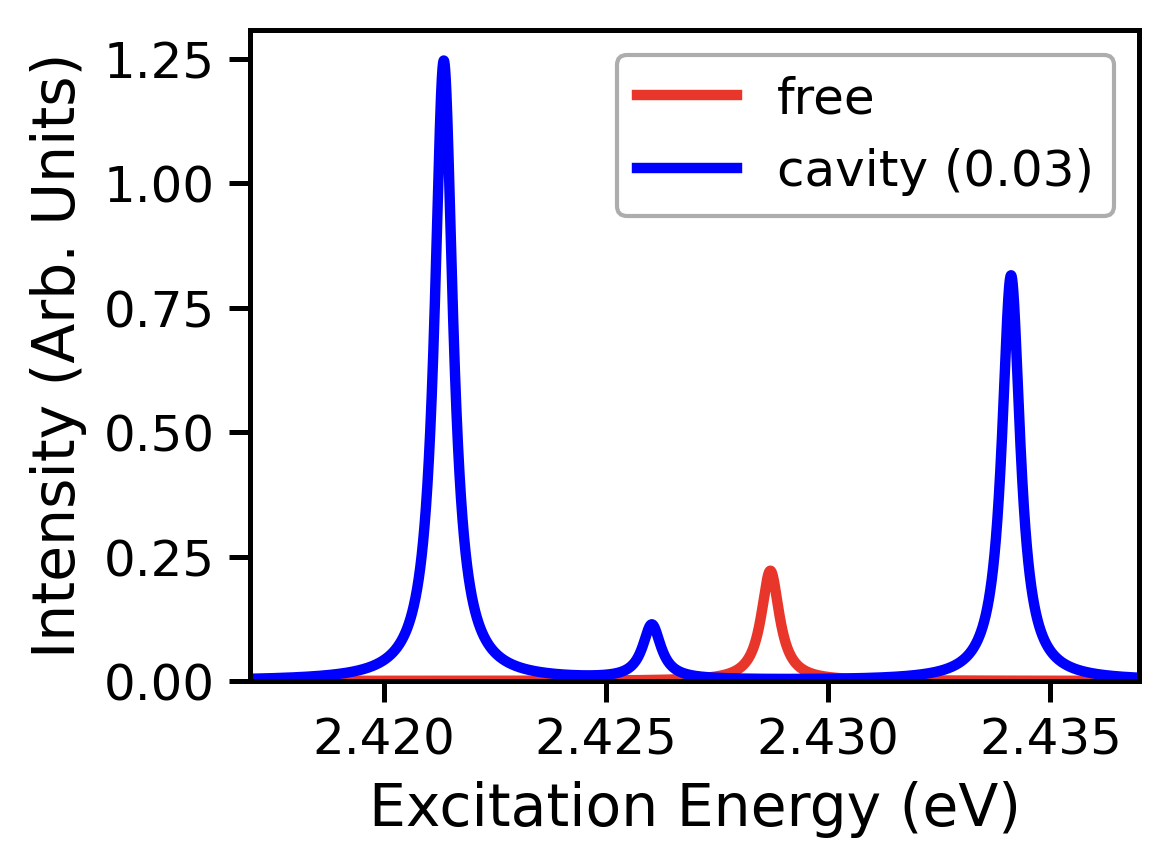} 
        \caption{}
        \label{fig:auh_0.03_1Pi_x}
    \end{subfigure}

    \caption{Absorption spectra of AuH inside an $x$-polarized optical cavity for coupling strengths of 0.01 and 0.03 a.u., with the cavity frequency tuned to resonance with the $\mathrm{1(I)}$ state arising from the $^3\Sigma^+$ manifold.}
    \label{fig:auh_0.01_vs_0.03}
\end{figure}

In the free molecule, the lowest excited states come from the $^3\Sigma^+$ manifold. Spin--orbit coupling splits this manifold into a dark $0^-$ state at 2.406~eV and a doubly degenerate
$1(\mathrm{I})$ state at 2.429~eV. In a non-relativistic picture, this transition is spin-forbidden. With spin--orbit coupling, the $1(\mathrm{I})$ state gains a small transition dipole, with
one component along $x$ and one along $y$. This gives the weak peak in the free spectrum (red curves). The bright, spin-allowed $0^+$ state lies much higher, at 2.879~eV, and is polarized along the bond, so it does not couple to the cavity. Figure~\ref{fig:auh_0.01_1Pi_x} shows the spectrum for $\lambda = 0.01$~a.u., with the cavity frequency set at 2.4313~eV. The $x$ component of the $1(\mathrm{I})$ state mixes with the cavity photon and forms a lower and an upper polariton, at 2.426 and 2.431~eV. The two states are almost equal mixtures of light and matter, and the Rabi splitting is 4.4~meV. The middle peak at 2.428~eV is the $y$ component of the $1(\mathrm{I})$ state. It cannot couple to the $x$-polarized field and stays near the position of free molecule's excitation energy. The two polaritons have very different intensities. The LP is about three times brighter than the free $x$ component, while the UP is almost dark, about forty times weaker than the LP. The reason is that the polaritons do not only share the small dipole of the triplet. Through the photon, they also borrow intensity from higher-lying $x$-polarized transitions of the molecule. At $\lambda = 0.03$~a.u.\ (Figure~\ref{fig:auh_0.03_1Pi_x}), with the cavity frequency set at 2.4572~eV, the Rabi splitting grows to 12.8~meV, about three times larger than at $\lambda = 0.01$~a.u., as expected for a splitting that grows with the coupling strength. The upper polariton, which was almost dark at $\lambda = 0.01$~a.u., is now clearly visible as well. The two peaks are about eleven and seven times brighter than the free $x$ component, and their total intensity is about six times larger than at $\lambda = 0.01$~a.u. At this coupling strength, the intensity that the polaritons borrow through the photon is much larger than the triplet's own transition dipole. A cavity can therefore couple strongly to a transition that is almost forbidden in the free molecule, and a stronger coupling turns this weak transition into two clearly visible peaks.

\subsection{Selenophene inside cavity}
As another example, we have studied selenophene (C$_4$H$_4$Se). In this molecule the heavy selenium atom gives a strong spin--orbit coupling, so it is a good test for the four-component treatment. We have placed the molecule in the $yz$ plane with the $C_2$ axis along $z$, and the cavity field is polarized along this axis. The cavity frequency is set to 0.19468~a.u. (5.297~eV) and the coupling strength is 0.01~a.u. The excited states have been computed with 4c-QED-ADC(2) on top of a QED-DHF reference. 
\begin{figure}[h!]
    \centering

    \begin{subfigure}{0.48\textwidth}
        \centering
        \includegraphics[width=\textwidth]{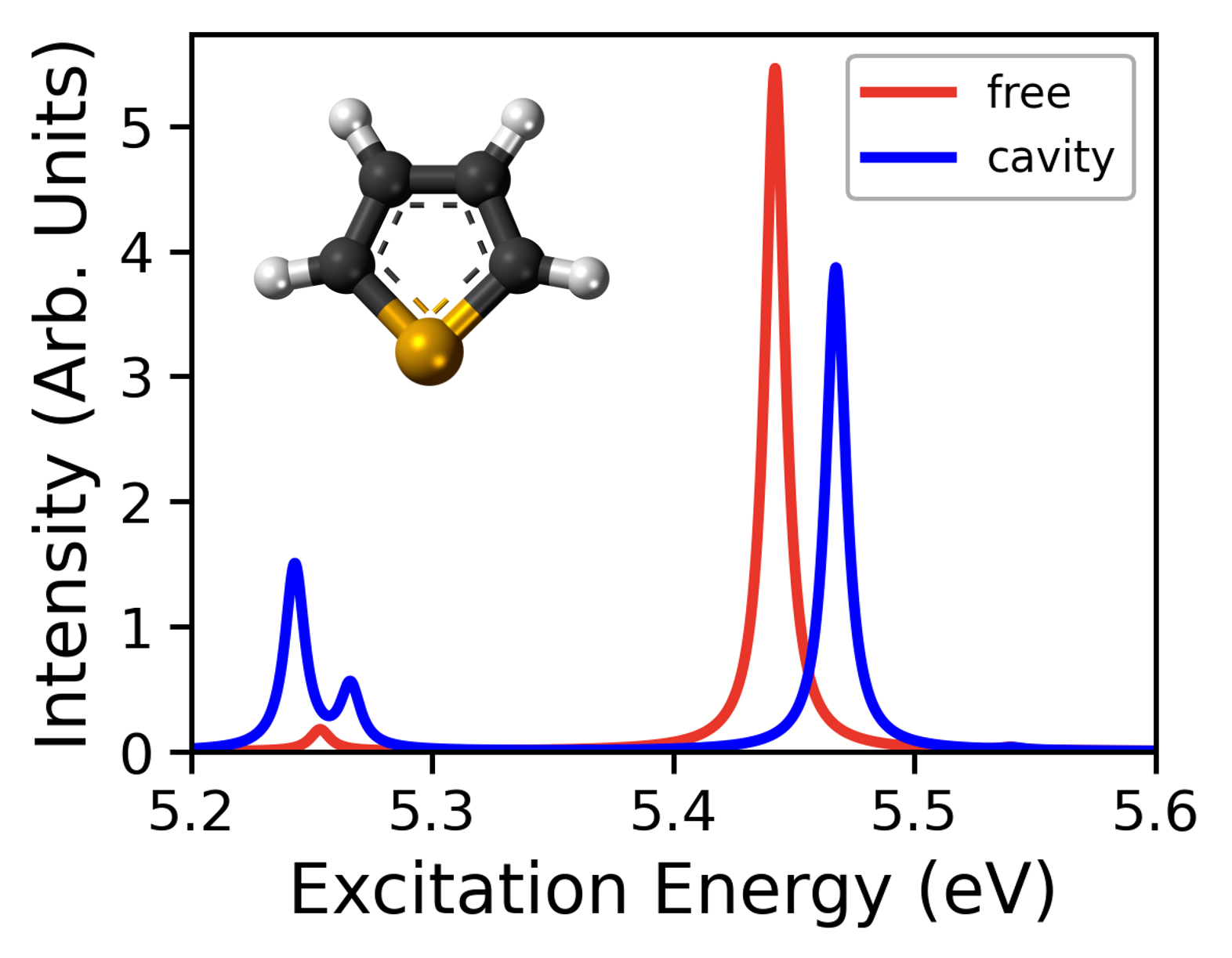}
        \caption{}
        \label{fig:selenophene_0.01_spectrum}
    \end{subfigure}

    \caption{Absorption spectra of selenophene outside (red) and inside (blue) a $z$-polarized
cavity with a coupling strength of 0.01~a.u. and cavity frequency of 5.2975~eV, tuned to resonance with the $1\,^3A_2$ state.}
    \label{fig:selenophene_0.01}
\end{figure}
The resulting absorption spectra are shown in Figure~\ref{fig:selenophene_0.01}, with a Lorentzian broadening of 0.01~eV. Without the cavity (free molecule, red curve), the spectrum in this region has two features. A strong peak at 5.442~eV comes from the bright, spin-allowed $2\,^1A_1$ state. A much weaker peak appears at 5.254~eV. It belongs to the $1\,^3A_2$ triplet state. In a non-relativistic picture, this transition is fully spin-forbidden and would not show up in the spectrum. With spin--orbit coupling, however, the triplet splits into three components, and one of them has a small $z$-polarized intensity. The other two components are polarized along $x$ and $y$ and are almost dark. In the cavity (blue curve), the spectrum changes clearly. The weak triplet peak is replaced by two peaks, at 5.243~eV and 5.266~eV. These are the lower and upper polaritons. They come from the mixing of the $z$-polarized triplet component with the cavity photon. At this frequency, the two states are almost equal mixtures of light and matter, with a Rabi splitting of 23~meV. This result shows that a spin-forbidden state can still
form polaritons, as long as spin--orbit coupling gives it a non-zero transition dipole. Such a state would stay completely uncoupled in a non-relativistic description. The $x$- and $y$-polarized triplet components do not couple to the $z$-polarized field, and they shift
by only about 1~meV. The resonance does not occur exactly at the bare triplet excitation energy. The cavity frequency that gives approximately equal light--matter mixing (5.2975~eV) lies about 44~meV above the free excitation energy (5.254~eV). This happens because the photon does not interact with
the triplet alone. It also couples strongly to the nearby bright $2\,^1A_1$ state, and the dipole self-energy shifts the matter states as well. Together, these effects move the point of best resonance away from the bare excitation energy. The polaritons are also much brighter than the free triplet state. The combined intensity of the LP and UP is about eleven times that of the uncoupled state. The LP alone is more than eight times brighter than the free triplet, and the UP is about three times brighter. At the same time, the bright $2\,^1A_1$ peak moves up by 25~meV, to 5.468~eV. It loses about 30\% of its intensity and gains about 11\% photon character. Therefore, the extra intensity of the polaritons is borrowed from the bright singlet state.

\subsection{Spin-Orbit Driven Reversal of Rabi Splitting}
\label{sec:hx_series}

Table~\ref{tab:so_splitting_hbr_hi} summarizes the Rabi splittings obtained for the HX (X = Br, I, At) molecules when the cavity frequency is tuned to the
\(^{3}\Pi_{0^{+}}\) and \(^{1}\Pi_{1}\) excitation energies of the free molecule, together with the corresponding cavity polarization directions. A clear and systematic trend emerges across the series as the halogen atom becomes heavier. For HBr, the Rabi splitting is larger when the cavity is tuned to the \(^{1}\Pi_{1}\) state than when it is tuned to the \(^{3}\Pi_{0^{+}}\) state. This behavior is consistent with spin-selection rules, since the singlet ground state couples more strongly to the singlet \(^{1}\Pi_{1}\) excited state, whereas the transition to the triplet \(^{3}\Pi_{0^{+}}\) state remains weak due to its predominantly triplet character. As the system moves to heavier halogens, spin--orbit coupling becomes increasingly important and qualitatively alters this picture.

\begin{table}[h!]
\centering
\caption{Rabi splittings \(\Omega\) (in meV) for HX (X = Br, I, At) molecules at a light–matter coupling strength of 0.01 a.u. The third column defines the excitation energies (in eV) of \(^{3}\Pi_{0^{+}}\) and \(^{1}\Pi_{1}\) of the free molecule, and $\hat{e}_{\alpha}$ signifies the directionality of the cavity mode.}
\label{tab:so_splitting_hbr_hi}
\begin{tabular*}{0.80\textwidth}{@{\extracolsep{\fill}}ccccr}
\hline
\hline
Molecule & State &\(\omega_{cav}\) &$\hat{e}_{\alpha}$  &\(\Omega\) \\
\hline
{HBr} 
 & \(^{3}\Pi_{0^{+}}\)   &6.7772 & $z$ & 17.7 \\
 & \(^{1}\Pi_{1}\)       &7.3162 & $x$ & 22.7 \\[2pt]
{HI}
 & \(^{3}\Pi_{0^{+}}\)   &5.6885 & $z$ & 33.7 \\
 & \(^{1}\Pi_{1}\)       &6.0945 & $x$ & 5.1 \\[2pt]
{HAt}
 & \(^{3}\Pi_{0^{+}}\)   &5.9555 & $z$ & 108.2 \\
 & \(^{1}\Pi_{1}\)       &5.3105 & $x$ & 10.8 \\
 \hline
 \hline
\end{tabular*}
\end{table}

For the heavier molecules, the picture changes. In HI, the splitting for the \(^{3}\Pi_{0^{+}}\) state (33.7~meV) is more than six times larger than that for the \(^{1}\Pi_{1}\) state (5.1~meV). In HAt, it grows to 108.2~meV, which is about six times the HBr value and ten times the \(^{1}\Pi_{1}\) splitting of the same molecule. The reason is spin--orbit coupling, which mixes states with the same $\Omega$. Spin--orbit coupling mixes the \(^{3}\Pi_{0^{+}}\) state with \(^{1}\Sigma^{+}\) states of the same symmetry ($0^{+}$, $z$-polarized), and the triplet borrows a sizeable transition dipole from them. This mixing becomes stronger as the halogen becomes heavier. The \(^{1}\Pi_{1}\) state, on the other hand, mixes with triplet states of $\Omega = 1$ and loses part of its singlet character. Its splitting drops from 22.7~meV in HBr to only 5.1 and 10.8~meV in HI and HAt, respectively. In HAt, the mixing is so strong that the state that is formally spin-forbidden couples to the cavity far more strongly than the formally allowed one. The ordering of the two splittings is therefore reversed in HBr relative to the heavier molecules. This reversal arises from spin–orbit coupling and cannot be captured by a nonrelativistic treatment.

\subsection{Modified Potential Energy Surface}

\begin{figure}[h!]
    \centering
    \begin{subfigure}{0.47\textwidth}
        \centering
        \includegraphics[width=\textwidth]{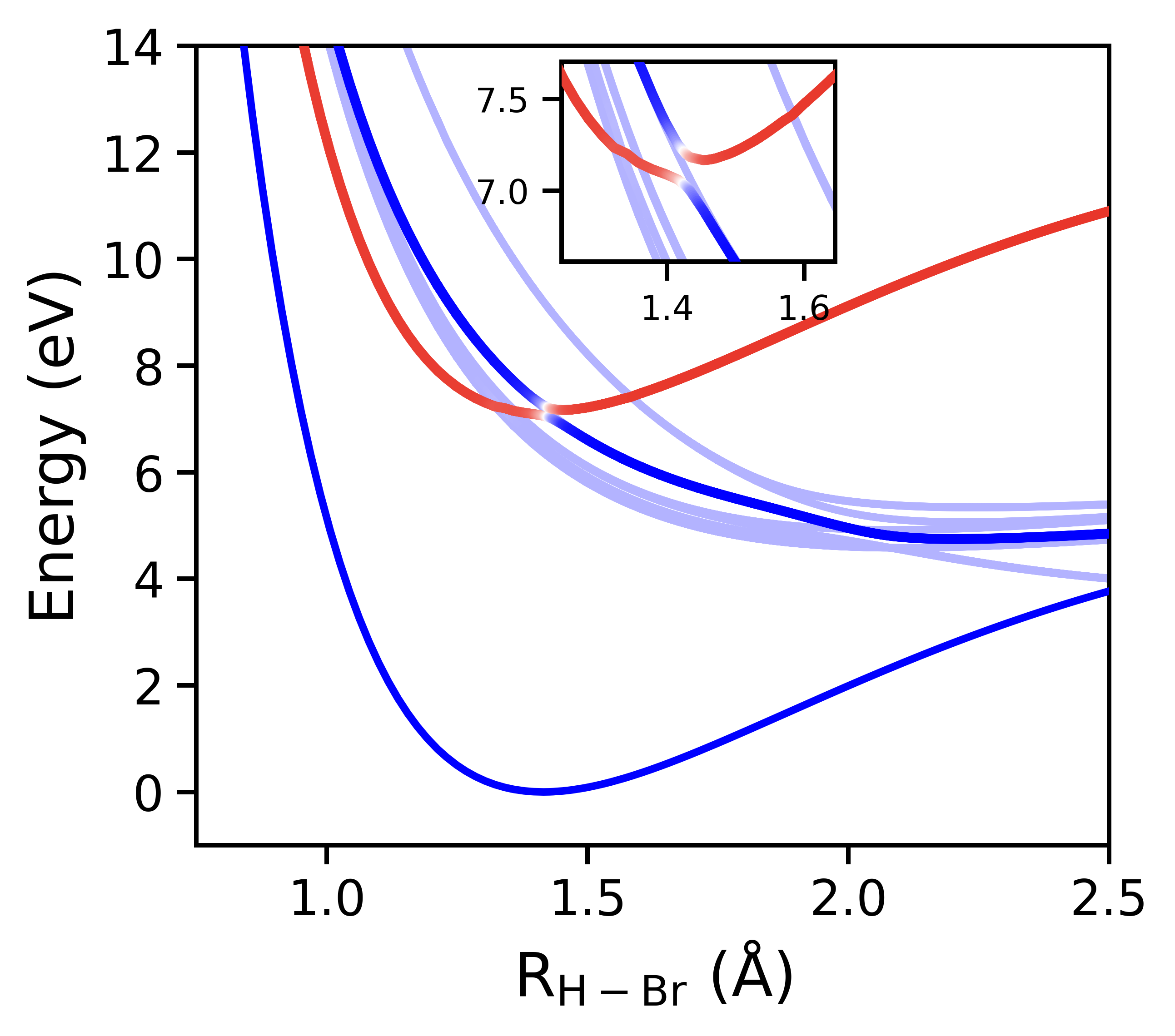}
        \caption{}
        \label{fig:hbr_1pi_x_0.05_pes}
    \end{subfigure}
    \begin{subfigure}{0.47\textwidth}
        \centering
        \includegraphics[width=\textwidth]{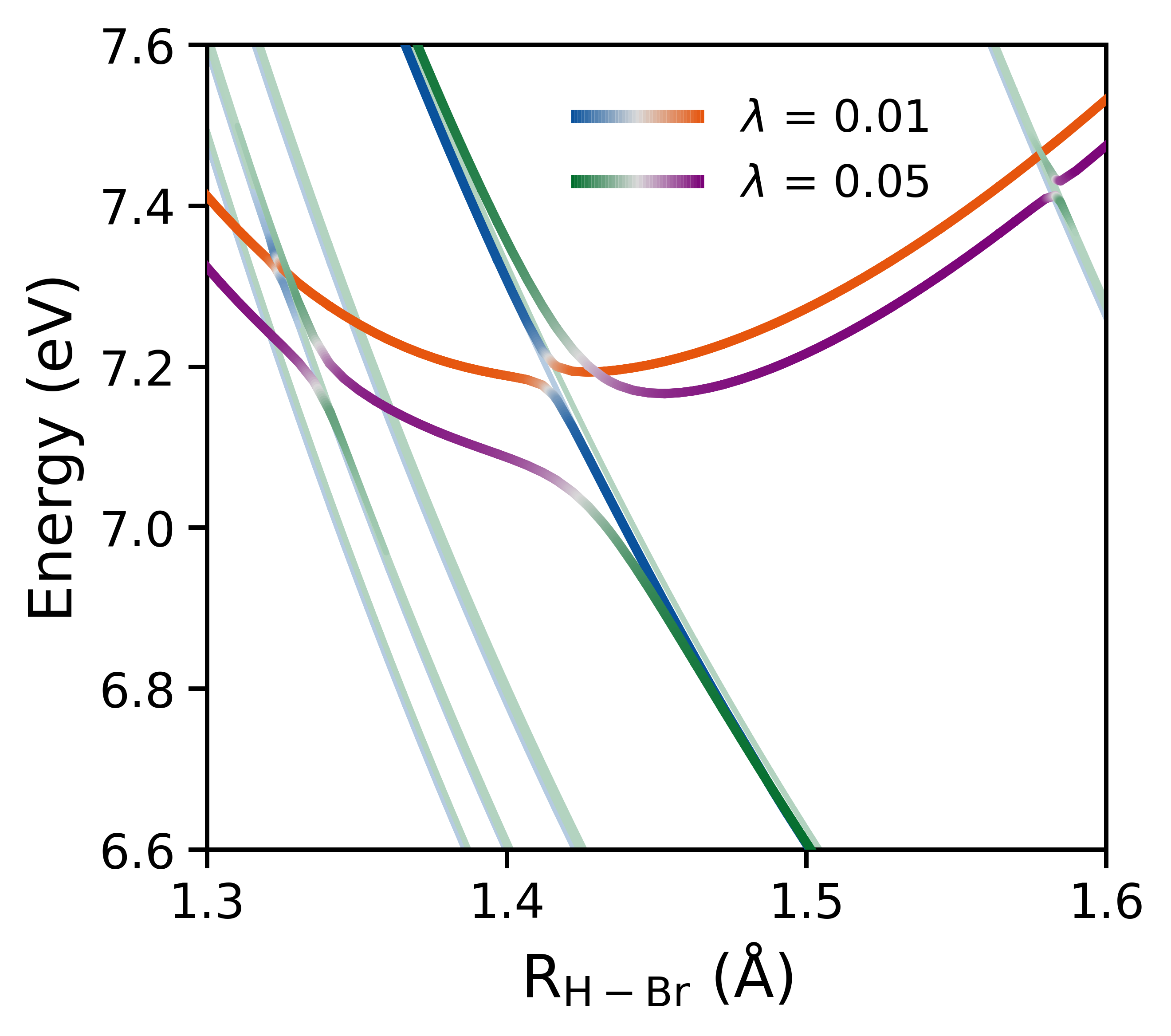}
        \caption{}
        \label{fig:hbr_1pi_zoom}
    \end{subfigure}

\caption{Potential energy curves of HBr in an optical cavity polarized along the
$x$-axis, with cavity frequency in resonance with the $^1\Pi$ state at the
equilibrium bond length. (a) All states for $\lambda = 0.05$~a.u. The colour
shows the character of each state, from electronic (blue) to photonic (red);
white marks an equal mixture. The inset zooms on the avoided crossing between
the lower and upper polaritons. (b) The avoided-crossing region for
$\lambda = 0.01$ and $0.05$~a.u. For $\lambda = 0.01$ the colour runs from
blue (electronic) to orange (photonic), and for $\lambda = 0.05$ from green
(electronic) to purple (photonic). Faint lines are the uncoupled states.}

    \label{fig:hbr_pes}
\end{figure}

We have investigated how a cavity mode in resonance with an electronic excitation modifies the potential energy curves of HBr. Figure~\ref{fig:hbr_pes}
shows the results for a cavity polarized along the $x$-axis, perpendicular to the H--Br bond. The cavity frequency, $\omega_c = 7.19$~eV, matches the $^1\Pi$ excitation energy of the free molecule at its equilibrium geometry. The colour of each curve shows its character. Blue for predominantly electronic, red for
predominantly photonic, and white for an approximately equal mixture. Figure~\ref{fig:hbr_1pi_x_0.05_pes} gives the PES for
$\lambda = 0.05$~a.u. The ground-state potential keeps its bound shape, with a minimum near 1.42~\AA. In the energy range of interest, the excited states come mainly from the $^3\Pi$ and $^1\Pi$ configurations. Their potentials are repulsive near the equilibrium geometry and fall towards the dissociation limit as the bond is stretched. In the uncoupled picture, the photonic state is the electronic ground state with one cavity photon, so its potential follows the
ground-state curve shifted upwards by $\omega_c$. 
The light--matter coupling between the cavity mode and the $^1\Pi$ electronic state turns into an avoided crossing and
produces a lower (LP) and an upper (UP) polaritonic branch. As shown in the inset of Figure~\ref{fig:hbr_1pi_x_0.05_pes}, the two branches come closest at $R = 1.43$~\AA, where their separation is about 0.17~eV. At this geometry, both LP and UP states are strongly mixed, with photonic contributions of about 50\%. The midpoint of the Rabi splitting between the hybrid states is 7.11~eV, which coincides with the energy of the uncoupled
$^1\Pi$ state at the same bond length, so the splitting is nearly symmetric about the electronic state. Across the crossing, the two branches exchange their character rapidly. Between 1.40 and 1.45~\AA, the photonic contribution to the LP falls from about 87\% to 10\%, and the UP changes in the opposite direction. As a result, the LP follows the photonic potential on the shorter bond-length region and the repulsive electronic surface on the longer bond-length region, reaching about 4.8~eV at 2.5~\AA. The UP follows the steep electronic character at short bond lengths and becomes mainly photonic as the bond is stretched. This exchange of character changes the shape of the surfaces. The UP develops a shallow minimum near 1.45~\AA{} at about 7.17~eV. The well forms because
the repulsive electronic state mixes with the bound photonic state, which gives bound character to a branch that would otherwise be repulsive. The LP, in contrast, connects a mainly photonic state at short bond lengths with a dissociative electronic state at long bond lengths. The cavity therefore creates
a bound region in the excited-state manifold. The interaction is also selective with respect to the transition-dipole direction, as seen in the previous sections. For an $x$-polarized cavity, only the component of the $^1\Pi$ pair with a transition dipole along $x$
couples to the photon mode. The orthogonal $y$ component stays essentially uncoupled and passes between the LP and UP through the avoided-crossing region and remains repulsive in nature.

Figure~\ref{fig:hbr_1pi_zoom} compares the avoided-crossing region for $\lambda = 0.01$ and $0.05$~a.u. Away from the crossing, the electronic parts of the branches are nearly identical for both couplings, so the coupling strength acts mainly near the crossing point. At $\lambda = 0.01$~a.u., the LP and UP come closest at 1.42~\AA{} with a gap of only about 39~meV. At $\lambda = 0.05$~a.u., the gap grows to about 175~meV. A five-fold increase in $\lambda$ thus increases the Rabi splitting by a factor of about 4.5. The stronger coupling
also widens the region over which the states mix. At $\lambda = 0.01$~a.u., the photonic contribution to the LP drops from 98\% to about 1\% between 1.40 and 1.43~\AA. At $\lambda = 0.05$~a.u., the same change is spread over a wider range of bond lengths, as the smooth colour gradient along the purple and green branches shows. The coupling strength therefore controls the shape of the upper polaritonic surface. At $\lambda = 0.01$~a.u., the UP minimum is very shallow (7.20~eV near $R = 1.43$~\AA) and hardly differs from the bare crossing point. At $\lambda = 0.05$~a.u., the minimum is lower (7.17~eV), lies at a longer bond length (1.45~\AA), and is clearly wider. The stronger coupling also lowers the photonic parts of both branches, the LP at short and the UP at long bond lengths, by about 0.06--0.09~eV relative to $\lambda = 0.01$~a.u., while the electronic parts are almost unchanged. This shift comes from the interaction of the photon mode with off-resonant electronic transitions. Although the cavity frequency is tuned to the $^1\Pi$ state, we see an extra polaritonic surface with a small avoided crossing near 1.34 Angstrom for a coupling strength of 0.05. It comes from the $^3\Pi$ manifold. Through spin-orbit coupling, one of its components borrows a small transition dipole from the neighboring $^1\Pi$ state, so it also couples to the cavity. This feature would not appear in a non-relativistic calculation, where the $^3\Pi$ states have zero transition dipole from the ground state. A similar feature is also observed for $\lambda = 0.01$~a.u., but it is too weak to be seen.

\subsection{Computation time}

\begin{figure}
    \centering
    \begin{subfigure}[b]{0.6\textwidth}
    \centering
    \includegraphics[width=\textwidth]{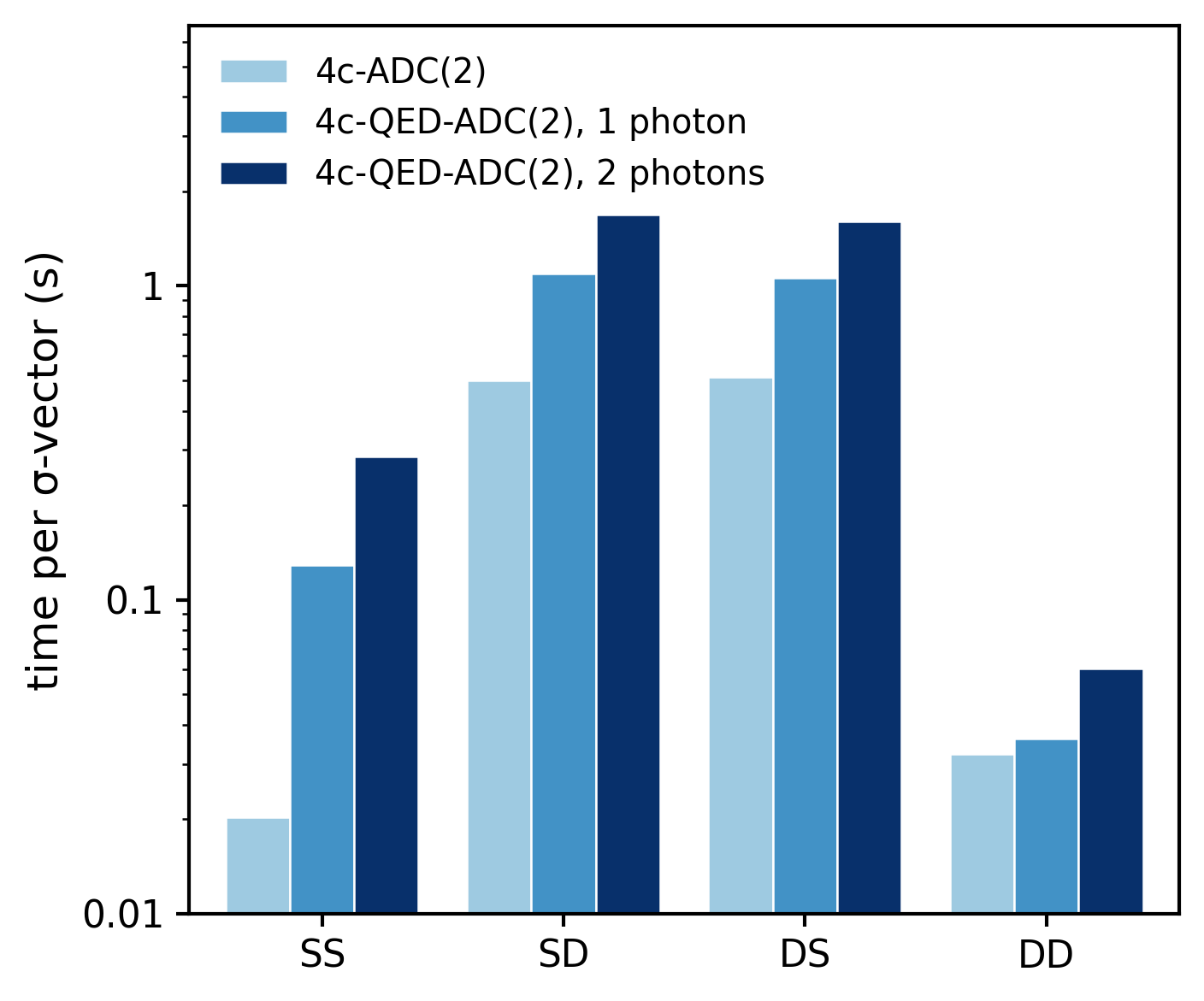}
    \end{subfigure}
    \caption{Wall times (in seconds) of the individual Hamiltonian blocks of 4c-QED-ADC(2) per $\sigma$-vector for AuH at a light--matter coupling strength of 0.02~a.u., comparing 1- and 2-photon dispersion with the free-molecule 4c-ADC(2) result.}
    \label{fig:time_calc}
\end{figure}

Figure~\ref{fig:time_calc} shows the wall time for one $\sigma$-vector evaluation for AuH, resolved by blocks of the 4c-QED-EE-ADC(2) matrix. The cavity calculations use a light--matter coupling strength of 0.02~a.u.\ and a cavity frequency of 3.19~eV. We compare the free-molecule 4c-EE-ADC(2) calculation with cavity calculations including up to one or two photons. Each calculation uses the dyall.v2z and unc-cc-pVDZ basis sets for Au and H, respectively, and a virtual-spinor correlation cutoff of 20~a.u. leading to a correlation space of 34 occupied and 114 virtual spinors. The ten lowest excited states are computed in each case. Adding the one- and two-photon sectors increases the configuration-space dimension from 3.6 million to 7.2 and 10.9 million, respectively. All calculations ran on a single shared-memory node with 314~GB of memory and two Intel\textsuperscript{\textregistered} Xeon\textsuperscript{\textregistered} Silver 4216 processors with a base clock frequency of 2.10 GHz.

The cost of each iteration of the Davidson solver is dominated by the matrix--vector product. For the free molecule, one $\sigma$-vector takes 1.07~s. With single- and double-photon dispersion, it takes 2.32~s and 3.64~s, i.e., 2.2 and 3.4 times longer. These factors are close to the growth of the configuration space. Each additional photon sector therefore costs about as much as one more electronic ADC(2) product, plus roughly 10--15\% for the coupling between the sectors.
Most of the time is spent in the SD and DS blocks, which couple the single (ph) and double (2p2h) excitations. They take 90--95\% of the $\sigma$-vector time in all three calculations. In the free molecule these blocks contain the usual ADC(2) coupling terms, which contract the doubles part of the vector with the two-electron integrals $\langle ja\|bc\rangle$ and $\langle jk\|ib\rangle$. For example, the SD block forms terms of the type $\sum_{jbc}\langle ja\|bc\rangle\, r_{ij}^{bc}$, and these contractions scale as $O(o^{2}v^{3})$. The DS block is the adjoint of the SD block. It uses the same integrals and the same contractions, only in the opposite direction, from the singles to the doubles. As a result, the two blocks do the same number of operations and take almost the same time, even though SD reads the large doubles vector and DS the much smaller singles vector. In the cavity calculations, two kinds of terms appear in these blocks. First, the electronic coupling described above is present in every photon sector, so it is repeated twice with single-photon dispersion and three times with double-photon dispersion. Each copy costs the same as in the free molecule, because these electronic terms do not depend on the photon number. Second, new terms appear that couple neighboring photon sectors. They come from the bilinear light--matter interaction and connect the singles of sector $n+1$ with the doubles of sector $n$ through the dipole integrals $\tilde{d}_{kc}$ as
\begin{align}
  \sigma^{(n+1)}_{ia} &\mathrel{+}= 2\sqrt{n+1}\,\sqrt{\tfrac{\omega}{2}}
      \sum_{kc} \tilde{d}_{kc}\, r^{(n)}_{ik,ac}, \\
  \sigma^{(n)}_{ij,ab} &\mathrel{+}= 2\sqrt{n+1}\,\sqrt{\tfrac{\omega}{2}}\,
      \mathcal{A}_{ij}\mathcal{A}_{ab}\bigl[\tilde{d}^{*}_{jb}\, r^{(n+1)}_{ia}\bigr],
\end{align}
where $\mathcal{A}_{ij}$ and $\mathcal{A}_{ab}$ antisymmetrize the indices, $\mathcal{A}_{ij}Z_{ij,ab} = \tfrac12\bigl(Z_{ij,ab}-Z_{ji,ab}\bigr)$. These terms only scale as $O(o^{2}v^{2})$, because each of them passes once over the doubles vector, so they are much cheaper than the electronic coupling. There is one such pair of sectors with single-photon dispersion and two with double-photon dispersion. With single-photon dispersion, the SD block takes 1.09~s, which is twice the free-molecule value (2~$\times$~0.50~s) plus 0.09~s for the photon coupling. With double-photon dispersion, it takes 1.69~s, which is three times the free-molecule value plus twice that coupling. The DS block behaves in the same way..

The SS block shows the fastest growth of all four blocks. Its cost rises from 20~ms for the free molecule to 129~ms with single-photon dispersion and 286~ms with double-photon dispersion, i.e., by factors of 6.4 and 14.1. This reflects that most of the new light--matter terms of the 4c-QED-ADC(2) matrix act within the singles space. For each photon sector, the singles block contains the electronic ADC(2) terms together with the photon energy $n\omega$ and a second-order light--matter correction. In addition, the singles of neighboring photon sectors are coupled through the dipole integrals $\tilde{d}_{ij}$ and $\tilde{d}_{ab}$ and second-order terms, and with double-photon dispersion a further block connects the zero- and two-photon sectors directly. The bare photon states and their coupling to the singles are also included in this block. Because the singles space is small, all of these parts are built once as dense matrices and are only multiplied with the trial vector in each iteration. The number of such matrix--vector products therefore grows from one for the free molecule to four with single-photon dispersion and to nine with double-photon dispersion. Even so, the SS block remains a minor part of the calculation, taking only 2\%, 6\%, and 8\% of the $\sigma$-vector time, and it scales as $O(o^{2}v^{2})$, compared with $O(o^{2}v^{3})$ for the SD and DS blocks. The DD block is cheap and diagonal at ADC(2) level. Each double amplitude is simply multiplied by its orbital-energy difference, shifted by $n\omega$ in photon sector $n$. For the free molecule, the SS block is the cheapest of the four, even cheaper than the diagonal DD block. In the cavity calculations, the picture is reversed. The SS block becomes 3.6 times more expensive than the DD block with single-photon dispersion and 4.7 times with double-photon dispersion. The reason is that the two blocks grow in different ways. The DD block gains only one diagonal per photon sector, because the photon modes do not couple the doubles of different sectors at this order. The SS block, in contrast, collects most of the light--matter coupling of 4c-QED-ADC(2). In addition to one singles block for each photon sector, it also holds the couplings between neighboring sectors and, when two photons are included, a direct coupling between the zero- and two-photon sectors.

\section{Conclusion}
In this work, we have presented the first fully relativistic 4c implementation of cavity quantum electrodynamical 4c-QED-EE-ADC(2) for the treatment of light–matter interactions. Two distinct variants of the 4c-QED-EE-ADC(2) approach have been realized. In the first variant, cavity effects are incorporated at the mean-field level through a modified Pauli–Fierz Hamiltonian, while in the second variant, cavity QED contributions are introduced exclusively at the relativistic electron-correlation level. These two types of implementation enable a systematic assessment of the role of cavity effects at different stages of the relativistic many-body treatment. 

To benchmark the methodology, we have computed cavity dispersion and electronic absorption spectra of the Hg atom confined within an optical cavity. These calculations demonstrate how both singlet–triplet and singlet–singlet electronic transitions are modified under strong light–matter coupling. Notably, even when the cavity mode is tuned into resonance with a singlet–triplet excitation, pronounced shifts and intensity changes are also observed for singlet–singlet transitions, because the cavity photon also couples off-resonantly to the bright S–S transition. With increasing light–matter coupling strength, state mixing becomes progressively more significant, leading to an enhancement of the otherwise weak singlet–triplet transition intensity inside the cavity. We have further demonstrated the applicability of the 4c-QED-ADC(2) framework to molecular systems, including AuH and halogen hydrides of the form HX (X = Br, I, At). For HX, the results have shown that with stronger relativistic effects from HBr to HAt, the Rabi splitting originally larger for the singlet excited state reverses and becomes larger for the triplet state, because of the spin--orbit coupling.
For the AuH molecule, the cavity mode was tuned to couple with the $0^{+}$ electronic state, and the evolution of the polaritonic spectral features was systematically analyzed as a function of the light–matter coupling strength. These results clearly illustrate the formation and tunability of polaritonic states in a relativistic molecular setting. We have also shown that spin-forbidden states can form polaritons when spin--orbit coupling gives them a small transition dipole. This is the case for the $1(\mathrm{I})$ state of AuH and for the $1\,^3A_2$ state of selenophene. These polaritons are much brighter than the free triplet state, because they borrow intensity through the photon from bright singlet transitions. For both Hg and AuH, we performed a comparative assessment of cavity effects introduced at different theoretical levels, namely with the cavity already included in the mean-field reference (4c-QED-DHF) and with the cavity included only at the correlated level on a cavity-free DHF reference. This comparison reveals that the non-polaritonic variant, yields excitation energies that closely track those obtained from the fully self-consistent 4c-QED-EE-ADC(2) approach at weak coupling. At $\lambda = 0.01$~a.u., the deviations are about 1~meV or less for the Rabi splittings and a few meV for the polariton positions. For AuH at $\lambda = 0.03$~a.u., the shift of the polariton pair below the free $0^+$ peak is about three times larger on the DHF reference than on the 4c-QED-DHF reference, and the dark states even shift in opposite directions on the two references. This behavior shows that the cavity parameters should be included already at the relativistic mean-field level when the coupling becomes strong. 
The effect of cavity coupling on the potential energy surfaces (PESs) of the HBr molecule was also investigated. Our results show that the qualitative nature of the ground-state PES remains essentially unchanged upon placing the molecule inside the cavity. In contrast, the excited-state PESs undergo significant modifications due to light–matter coupling. Specifically, the upper polaritonic surface acquires a bound character, while the lower polaritonic surface changes from mainly photonic character at short bond lengths to the repulsive electronic surface of the free molecule at longer bond lengths. The depth and width of the well on the upper surface grow with the coupling strength. At $\lambda = 0.05$~a.u., a component of the $^3\Pi$ manifold also forms a small avoided crossing with the photon, through the transition dipole it gains from spin--orbit coupling, which a non-relativistic treatment would miss.

In addition, we have quantified the computational overhead introduced by the inclusion of photonic degrees of freedom in the 4c-QED-EE-ADC(2) Hamiltonian. The enlarged excitation space and the increased cost of evaluating matrix–vector products make one $\sigma$-vector evaluation for AuH about 2.2 and 3.4 times more expensive than in the corresponding standard 4c-EE-ADC(2) calculation, with single- and double-photon dispersion, respectively. These factors follow the growth of the configuration space, and most of the time is spent in the coupling between single and double excitations. The associated computational cost also motivates the development of low-cost and efficient implementations. Ongoing work aims to improve the efficiency of the present framework and extend its application to larger, more complex systems.

\section{Acknowledgments}
AKD, SC, and AM acknowledge the support from IIT Bombay, ANRF (Project No. CRG/2023/00255), and ISRO for financial support, the IIT Bombay supercomputing facility, and C-DAC. Supercomputing resources (Param Smriti and Param Brahma) for computational time.
AKD acknowledges the research fellowship funded by the EU's NextGenerationEU through the Recovery and Resilience Plan for Slovakia, under project No.~09I03-03-V04-00117.
LK acknowledges funding from the EU's NextGenerationEU through the Recovery and Resilience Plan for Slovakia, under project No.~09I05-03-V02-00034, VEGA (grant No.~1/0670/24) and
the IMPULZ program of the Slovak Academy of Sciences under the Agreement on the Provision of Funds No. IM-2025-182.
SC acknowledges the Prime Minister's Research Fellowship (PMRF).
AM acknowledges CSIR SRF Fellowship.

\bibliography{ref}

\end{document}